\documentclass[twocolumn,trackchanges,resetfootnote]{aastex7}

\renewcommand{\thefootnote}{\dag\arabic{footnote}}

\shorttitle{Environments of quiescent galaxies at $1<z<5$}
\shortauthors{Kakimoto et al.}
\DeclareUnicodeCharacter{2061}{}
\begin{document}

\title{The role of small-scale environments in the quenching of massive galaxies at $1<z<5$}

\correspondingauthor{Takumi Kakimoto}

\author[0000-0003-2918-9890]{Takumi Kakimoto}
\affiliation{Department of Astronomical Science, The Graduate University for Advanced Studies, SOKENDAI, 2-21-1 Osawa, Mitaka, Tokyo 181-8588, Japan}
\affiliation{National Astronomical Observatory of Japan, 2-21-1 Osawa, Mitaka, Tokyo 181-8588, Japan}
\email[show]{takumi.kakimoto@grad.nao.ac.jp}  

\author[0000-0002-5011-5178]{Masayuki Tanaka}
\affiliation{National Astronomical Observatory of Japan, 2-21-1 Osawa, Mitaka, Tokyo 181-8588, Japan}
\affiliation{Department of Astronomical Science, The Graduate University for Advanced Studies, SOKENDAI, 2-21-1 Osawa, Mitaka, Tokyo 181-8588, Japan}
\email{masayuki.tanaka@nao.ac.jp}

\author[0000-0002-4225-4477]{Makoto Ando}
\affiliation{Institute for Cosmic Ray Research, The University of Tokyo, 5-1-5 Kashiwanoha, Kashiwa, Chiba 277-8582, Japan}
\affiliation{National Astronomical Observatory of Japan, 2-21-1 Osawa, Mitaka, Tokyo 181-8588, Japan}
\email{makoto.ando.astro@gmail.com} 

\author[0000-0002-9453-0381]{Kei Ito}
\affiliation{Cosmic Dawn Center (DAWN), Denmark}
\affiliation{DTU Space, Technical University of Denmark, Elektrovej 327, DK2800 Kgs. Lyngby, Denmark}
\email{keiit@dtu.dk} 

\author[0000-0003-4985-0201]{Ken Mawatari}
\affil{Waseda Research Institute for Science and Engineering, Faculty of Science and Engineering, Waseda University, 3-4-1 Okubo, Shinjuku, Tokyo 169-8555, Japan}
\affiliation{Department of Pure and Applied Physics, School of Advanced Science and Engineering, Faculty of Science and Engineering, Waseda University, 3-4-1 Okubo, Shinjuku, Tokyo 169-8555, Japan}
\email{mawatari@aoni.waseda.jp} 

\author[0000-0003-3228-7264]{Masato Onodera}
\affiliation{Subaru Telescope, National Astronomical Observatory of Japan, National Institutes of Natural Sciences (NINS), 650 North A'ohoku Place, Hilo, HI 96720, USA}
\affiliation{Department of Astronomical Science, The Graduate University for Advanced Studies, SOKENDAI, 2-21-1 Osawa, Mitaka, Tokyo 181-8588, Japan}
\email{monodera@naoj.org} 

\author[0000-0003-4442-2750]{Rhythm Shimakawa}
\affiliation{Waseda Institute for Advanced Study (WIAS), Waseda University, 1-21-1, Nishi-Waseda, Shinjuku, Tokyo 169-0051, Japan}
\email{rhythm.shimakawa@aoni.waseda.jp}

\author[0000-0001-6477-4011]{Francesco Valentino}
\affiliation{Cosmic Dawn Center (DAWN), Denmark}
\affiliation{DTU Space, Technical University of Denmark, Elektrovej 327, DK2800 Kgs. Lyngby, Denmark}
\email{fmava@dtu.dk} 

\author[0000-0002-9665-0440]{Po-Feng Wu}
\affiliation{Graduate Institute of Astrophysics, National Taiwan University, Taipei 10617, Taiwan}
\affiliation{Department of Physics and Center for Theoretical Physics, National Taiwan University, Taipei 10617, Taiwan}
\affiliation{Physics Division, National Center for Theoretical Sciences, Taipei 10617, Taiwan}
\email{wupofeng@phys.ntu.edu.tw} 

\author[0000-0001-6229-4858]{Kiyoto Yabe}
\affiliation{Subaru Telescope, National Astronomical Observatory of Japan, National Institutes of Natural Sciences (NINS), 650 North A'ohoku Place, Hilo, HI 96720, USA}
\affiliation{Department of Astronomical Science, The Graduate University for Advanced Studies, SOKENDAI, 2-21-1 Osawa, Mitaka, Tokyo 181-8588, Japan}
\email{kiyoyabe@naoj.org}

\author[0000-0002-8412-7951]{Shuowen Jin}
\affiliation{Cosmic Dawn Center (DAWN), Denmark}
\affiliation{DTU Space, Technical University of Denmark, Elektrovej 327, DK2800 Kgs. Lyngby, Denmark}
\email{shuji@space.dtu.dk} 

\author[0000-0002-7598-5292]{Mariko Kubo}
\affiliation{Department of Physics and Astronomy, School of Science, Kwansei Gakuin University, 1 Gakuen Uegahara, Sanda, Hyogo 669-1330, Japan}
\affiliation{Astronomical Institute, Tohoku University, 6-3, Aramaki, Aoba, Sendai, Miyagi 980-8578, Japan}
\email{markubo@kwansei.ac.jp} 

\author[0000-0003-3631-7176]{Sune Toft}
\affiliation{Cosmic Dawn Center (DAWN), Denmark}
\affiliation{Niels Bohr Institute, University of Copenhagen, Jagtvej 128, DK2200 Copenhagen N, Denmark}
\email{sune@nbi.ku.dk} 



\begin{abstract}
Massive quiescent galaxies (QGs) at high redshifts are likely progenitors of massive elliptical galaxies in the local Universe. Recent discoveries of high-redshift QGs in overdensities (galaxy groups and proto-clusters) have highlighted the relationship between massive galaxy quenching and their surrounding environment. We spectroscopically confirm a galaxy group associated with a massive QG at $z_\mathrm{spec}=4.53$ from the Lyman break feature using Subaru/FOCAS. This group consists of at least three star-forming galaxies within 150\,pkpc of the QG, suggesting a physical association necessary for galaxy quenching.
To investigate the role of the surrounding environment, we also perform a statistical analysis to characterize the typical environment of QGs at high redshifts. By selecting QGs using an SFR-based selection in the COSMOS field, we find that the quiescent fraction is higher in overdensities than in the field at $1<z<5$. In particular, the elevated quiescent fraction within small-scale overdensities ($\sim 200\,\mathrm{pkpc}$) demonstrates that environmental quenching, possibly driven by galaxy mergers and interactions, plays a major role in the evolution of massive QGs at high redshifts. By extending the low-redshift mass and environmental quenching model up to $z \sim 5$, we confirm similar trends to those inferred from the SDSS red galaxy fraction at $z<1$. While environmental dependence weakens at $z \gtrsim 3$ for the overall population, a clear correlation persists specifically for more massive galaxies. This mass-dependent environmental effect suggests a downsizing trend, highlighting that the accelerated evolution of massive galaxies in high-density regions is already in place in the early Universe.
\end{abstract}



\section{Introduction} \label{sec:intro}
Massive elliptical galaxies in the local Universe are known to be dominated by old stellar populations with little cold gas and dust, exhibiting no active ongoing star formation activity \citep[e.g.,][]{1991ARA&A..29..581Y,Gallazzi_2005,Bellstedt2020}. Since these massive galaxies likely have played a major role in cosmic star formation activities in the early Universe, it is crucial to understand their formation mechanisms. The star formation histories (SFHs) of massive elliptical galaxies, inferred from detailed spectroscopic observations at low redshifts, indicate that they experienced an intense starburst in the early Universe and then rapidly quenched \citep[stopped their star formation;][]{Nelan_2005,Thomas_2005,Thomas2010,Renzini_2006}. However, the physical drivers of this starburst and subsequent quenching remain unclear, representing one of the most outstanding issues in the field of galaxy formation and evolution \citep{Man2018}.

In the local Universe, it is well established that galaxy properties are strongly correlated with their environment and stellar mass \citep[e.g.,][]{1980ApJ...236..351D, Bower1992, Blanton2007, Peng_2010}. Consequently, the relative roles of internal ``mass quenching'' and external ``environmental quenching'' processes have been extensively discussed in recent literature \citep[e.g.,][]{Man2018,Donnari2021,Alberts2022,Lucia2025}. However, the relationship between galaxy colors (or physical properties) and the surrounding environment evolves with redshift. For instance, \citet{Butcher1984} show that galaxy clusters at higher redshifts $(0.3 < z < 0.5)$ have a higher fraction of blue galaxies (Butcher--Oemler effect), meaning spiral galaxies are more prevalent in galaxy clusters at these redshifts. Furthermore, previous studies at $z>1$ suggest the observed decrease in environmental quenching efficiency \citep[e.g.,][]{Nantais2016, Edward2024}.

Despite these environmental trends, massive elliptical galaxies in high-density environment are thought to have formed in the early Universe \citep[roughly at $3<z<5$;][]{Thomas_2005, Thomas2010} and are suggested to be old even at high redshifts \citep{Tanaka2024}. Therefore, it is necessary to directly observe their formation epochs to examine the formation process. At $z > 2$, previous studies have confirmed the existence of ``proto-clusters'', which are overdensity regions of galaxies considered to be the progenitors of local clusters \citep[][for a review]{Overzier_2016}. Spectroscopic confirmation of such proto-cluster populations indicates that they contain (dusty) star-forming and/or star-bursting galaxies, especially in the early Universe \citep[e.g.,][]{Miller2018,Mitsuhashi2021,Lemaux2022,Shah2024}. Interestingly, however, proto-clusters with several massive quiescent galaxies (QGs) have also been discovered.

Recently, \citet{Kakimoto2024} confirmed a massive QG at $z=4.53$. The galaxy is identified a young post-starburst galaxy located in a group environment, suggesting that some quenching processes are regulated by environmental effects such as mergers and/or interactions. In particular, a massive companion galaxy at 13\,pkpc\footnote{Hereafter, the prefix `p' indicates proper distances (e.g., pMpc, pkpc).} from the QG is likely physical associated although the spectroscopic confirmation is needed. Gas depletion due to a merger-induced starburst and/or feedback from a merger-induced AGN \citep[e.g.,][]{Springel2005,Hopkins2006,Hopkins2008,Dubois2016} may be responsible for the quenching. This case hints that the group environment may indeed be important for quenching even in the early Universe. However, we have thus far studied only a handful of such systems \citep[e.g.,][]{Jekyll2018,Kubo_2021,Kalita_2021,McConachie_2022,Ito_2023,Tanaka2024,Jin2024,Baker2026}, and it is essential to statistically extend the investigation of the relation between QGs and structure formation.

In this paper, we report on the spectroscopic confirmation of the dense group at $z=4.5$ and explore the environmental dependence of QGs at $1<z<5$. We define the galaxy environment using galaxy surface number density at each redshift slice, and examine the QG fraction as a function of the galaxy density. This paper is structured as follows. First, we describe and discuss the spectroscopic confirmation of the group with a QG at $z=4.53$ as one of the case studies in Section \ref{sec:453}. Then, we introduce the data and sample selection in Section \ref{sec:obs} for our statistical study of the environmental dependence. In Section \ref{sec:SEDfit}, the correlation between QG fraction and galaxy environment or stellar mass is summarized. Then, we discuss possible interpretation for evolution of QGs at high redshifts in Section \ref{sec:group}. Finally, we conclude the paper in Section \ref{sec:Concl}. We assume a \citet{Chabrier_2003} initial mass function (IMF) and a flat $\mathrm{\Lambda CDM}$ cosmology with $H_0 = 70\,\mathrm{km\,s^{-1}\,Mpc^{-1}}, \Omega_m = 0.3$, and $\Omega_\Lambda = 0.7$. All magnitudes are in the AB system \citep{1983ApJ...266..713O}. 

\section{Spectroscopic confirmation of a dense group at $z=4.5$} \label{sec:453}
\begin{figure*}
    \centering
    \includegraphics[width=0.9\textwidth]{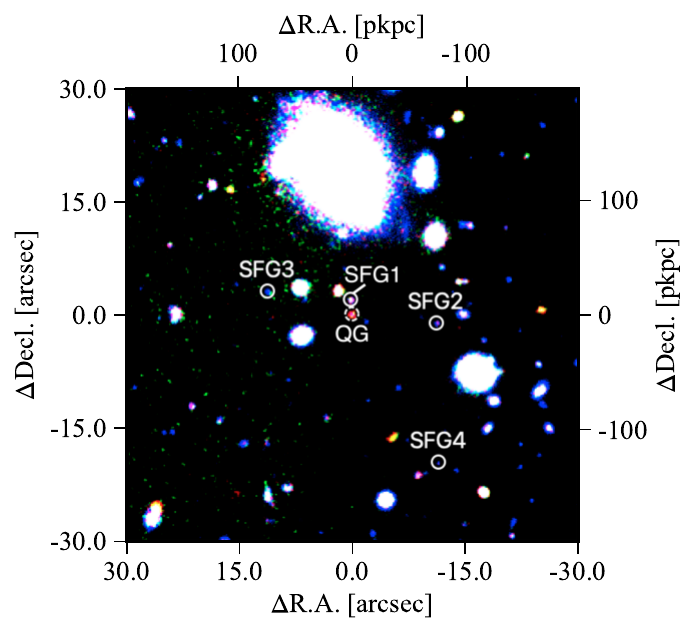}
    \caption{A pseudo-color image of the area around the QG (COSMOS-1047519) at $z_\mathrm{spec}=4.53$ (Red: VIRCAM/$K_s$-band, Green: VIRCAM/$H$-band, Blue: HSC/$i$-band; \citealp{McCracken_2012,Tanaka_2022}).}
    \label{fig:dist}
\end{figure*}
First, we briefly introduce the spectroscopic follow-up observation of candidate members of a galaxy group with a QG at $z=4.53$ \citep[COSMOS-1047519;][]{Kakimoto2024}. 

\subsection{COSMOS2020} \label{subsec:target}
\begin{deluxetable*}{llll}
\tablecaption{Summary of COSMOS2020 \label{tab:surveys}}
\tablehead{
\colhead{Survey name} & \colhead{Telescope/Instrument name} & \colhead{Band name} & \colhead{References} 
}
\colnumbers
\startdata
CLAUDS survey & CFHT/MegaCam & $u^*, u$ & \citet{Sawicki2019}\\ 
\parbox[t]{3cm}{Subaru COSMOS 20 project} & Subaru/Suprime-Cam & \parbox[t]{4.2cm}{$B,V,g,r,i,z,y,$ IB427, IB464, IB484, IB505, IB527, IB574, IB624, IB679, IB709, IB738, IB767, IB827, NB711, NB816} & \citet{Taniguchi2015} \\ 
HSC-SSP PDR2 & Subaru/Hyper Suprime-Cam & $g,r,i,z,y$ & \citet{Aihara2019} \\ 
\parbox[t]{3cm}{UltraVISTA survey DR4} & VISTA/VIRCAM & $Y,J,H,K_s,$ NB811 & \parbox[t]{4.5cm}{\citet{McCracken_2012}; \citet{NB1882013}; \citet{Moneti2023}} \\ 
Cosmic Dawn survey & Spitzer/IRAC & Ch1, Ch2, Ch3, Ch4 & \parbox[t]{4.5cm}{\citet{Ashby2013,Ashby2015,Ashby2018}; \citet{Steinhardt2014}; \citet{DAWN2022}} \\ 
\enddata

\end{deluxetable*}

The Cosmic Evolution Survey \citep[COSMOS;][]{Scoville2007} is a $2\, \mathrm{deg^2}$ survey to address the formation and evolution of galaxies. This is one of the widest fields of multi-band wavelength coverage to probe a large sample of galaxies and is centered at $\mathrm{R.A.=10^h00^m28^s.600, Decl.=+02^\circ 12'21''.000}$. We utilize COSMOS2020 \citep{Weaver_2022}, which consists of $\sim 970,000$ sources measured from the UV to the IR. The photometry is obtained from more than 30 filters, summarized in Table \ref{tab:surveys}.
The COSMOS2020 ``\textsc{Classic}'' catalog is utilized, whose photometry extracted with \texttt{SExtractor} \citep{Bertin1996} for the optical/NIR and with \texttt{IRACLEAN} \citep{IRACLEAN2012} for the IR. After removing contamination with bright stars from HSC mask and choosing the region observed with the VISTA and Suprime-Cam, we have a $1.279\,\mathrm{deg^2}$ area in COSMOS \citep{Weaver_2022complete}. We note that the group presented herein is outside the COSMOS-Web \citep{Casey2023}.

\subsection{Observation and target spectrum} \label{subsec:method}
\begin{figure*}[tb]
    \centering
    \begin{minipage}{0.49\textwidth}
        \centering
        \includegraphics[width=0.95\textwidth]{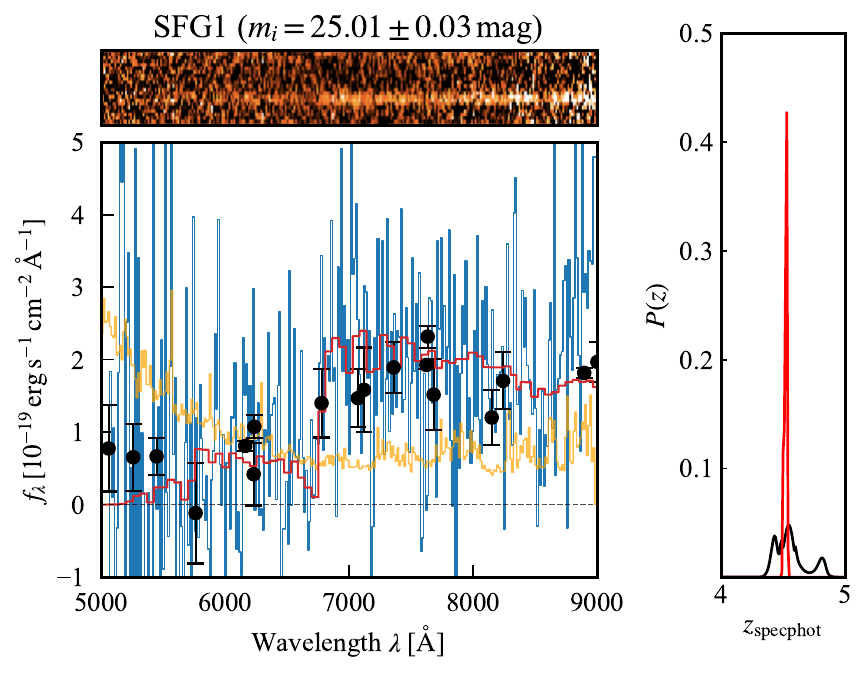}
    \end{minipage}
    \begin{minipage}{0.49\textwidth}
        \centering
        \includegraphics[width=0.95\textwidth]{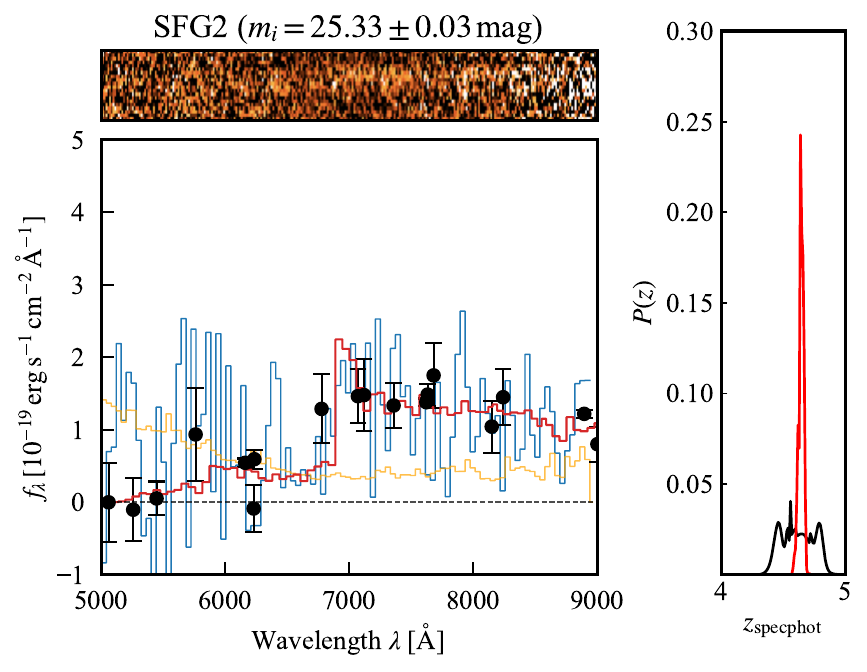}
    \end{minipage}
    \begin{minipage}{0.49\textwidth}
        \centering
        \includegraphics[width=0.95\textwidth]{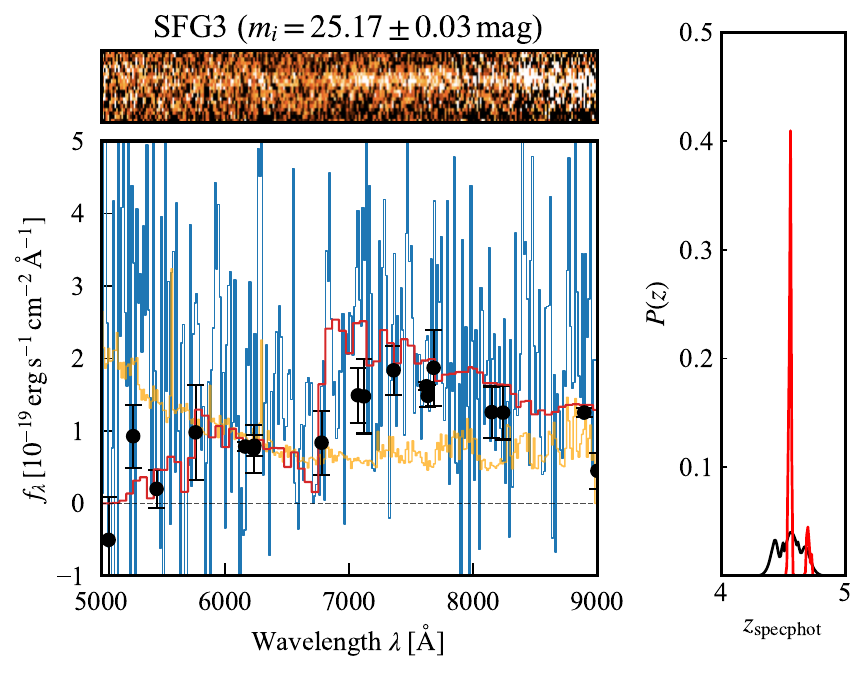}
    \end{minipage}
    \begin{minipage}{0.49\textwidth}
        \centering
        \includegraphics[width=0.95\textwidth]{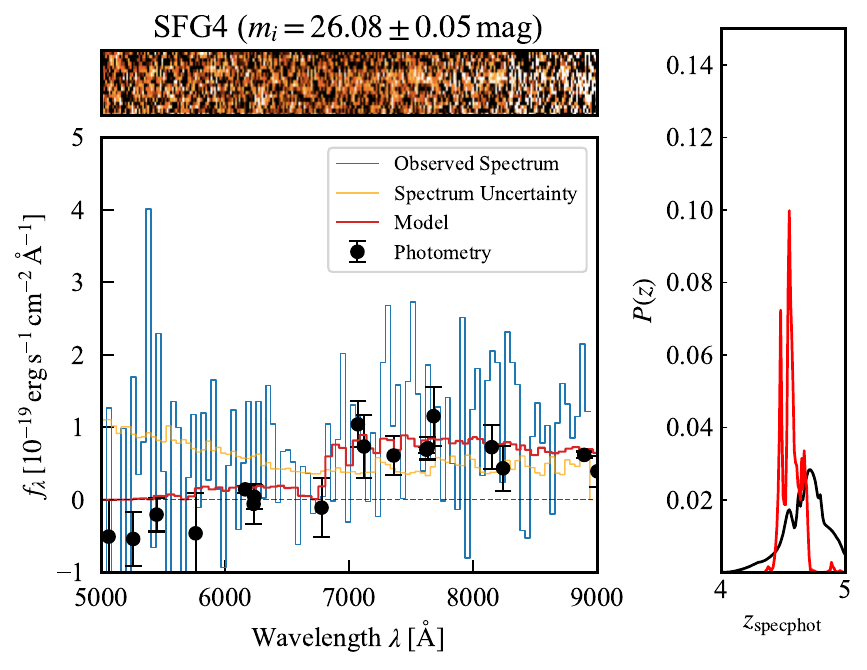}
    \end{minipage}
    \caption{Observed spectra of group members with Subaru/FOCAS (the blue line), spectral uncertainty (the yellow line), best-fitted SED from \texttt{prospector} (the red line), and photometry from COSMOS2020 (the black points). The spatial pixels (vertical axes) of 2D spectra on the upper panel are rescaled with the factor of 2. The right panels show the probability distribution of photometric redshift (the black line) and spectrophotometric redshift inferred from \texttt{prospector} (the red line).}
    \label{fig:Specgroup}
\end{figure*}

The overdenisty around the QG (COSMOS-1047519) is based on the photometric redshift estimates inferred from the COSMOS2020 catalog, and a follow-up spectroscopic observation is necessary to confirm the target and physical association of the members. In addition to the QG at the center, this group has four other candidate star-forming members (COSMOS-1048346, 1046859, 1048201, and 1041229 hereafter ``SFG1'', ``SFG2'', ``SFG3'', and ``SFG4'', respectively) with UV luminous SEDs (Figure \ref{fig:dist}). We attempt to constrain their spectroscopic redshifts from the Lyman break feature using optical light spectrograph.

Spectroscopic confirmation of massive star-forming galaxies (SFGs) at high redshifts can be challenging. The Ly$\alpha$ emission, which is the most luminous emission line from SFGs, is easily scattered by neutral hydrogen and obscured by surrounding dust. Especially at high redshift, while some galaxies have strong Ly$\alpha$ emission (called Ly$\alpha$ Emitter: LAE), UV luminous (massive) SFGs tend to show no Ly$\alpha$ emission \citep[e.g.,][]{Ando2006, Lai2008, Ono2010, Stark2010, Shimakawa2017, Weiss2021}. They instead exhibit a continuum break around Ly$\alpha$ and they are classified as Lyman Break Galaxies (LBG). Their UV continuum originates from star formation activity (massive stars), and the flux discontinuity is due to attenuation by neutral hydrogen, which can be used to constrain redshifts. Based on the Ly$\alpha$ fraction as a function of UV magnitude from \citet{Stark2010}, we expected that 10--20\% of our targets would show Ly$\alpha$ emission. In order to securely constrain their redshifts spectroscopically, we set the exposure time sufficient to detect the UV continuum.

Spectroscopic follow-up is conducted by the Faint Object Camera and Spectrograph \citep[FOCAS;][]{Kashikawa2002} on the Subaru telescope. Multi-object spectrograph mode is used to confirm all targets with only one mask. To detect the Ly$\alpha$ emission and/or Lyman break feature, 300B+SY47 Grism/filter is used, which covered a wavelength region of 4700--9100\,\AA\ with a resolution of $R\sim 1000$. Observation was conducted in March 2024 with a half night ($\sim 4$ hours on source exposure). We conduct the common spectral data reduction procedures (flat fielding, wavelength calibration, and background subtraction) with IRAF package \citep{Tody1986, Tody1993}. Summing up all exposures, flux calibration, and 1D spectrum extraction are operated by the same methods in \citet{Kakimoto2024}. The flux and FWHM of a bright star in each exposure is used to evaluate sky conditions. The flux calibration is based on the comparison between observed spectra and HST/CALSPEC \citep{Bohlin_2014} spectrum of GD71 (a white dwarf spectrophotometric standard star). Optimal extraction \citep{Horne1986} is performed for deriving 1D spectrum of each target \citep[see][in detail]{Kakimoto2024}.

Figure \ref{fig:Specgroup} shows the FOCAS spectra of the group members. The spectra are re-binned over 10 pixels for SFG1 and SFG3; for SFG2 and SFG4, a larger binning of 30 pixels is applied to gain the S/N by clipping the upper and lower 10\% of flux pixels, thereby excluding sky residuals. All SFGs do not show the strong Ly$\alpha$ emission, but a continuum flux at wavelengths longer than $\sim 6700$\,\AA\ that is 2--3 times brighter than at shorter wavelengths. For instance, the median S/N values at 6800--7500\,\AA\ are 3.33 for SFG1, 3.75 for SFG3 (per 13.4\,\AA), 4.25 for SFG2, and 2.41 for SFG4 (per 40.2\,\AA).

\subsection{Redshift estimation} \label{subsec:spec}
In order to constrain their redshift, we conduct SED fitting analysis with both FOCAS spectra, which trace the Lyman break with a high spectral resolution, and photometry from COSMOS2020 covering a wider wavelength range with a lower spectral resolution using \texttt{prospector} \citep{Johnson_2021}. We use the MILES stellar library \citep{Falcon-Barroso2011} and MIST isochrones \citep{Choi2016,Dotter2016}. We further assume a \citet{Chabrier_2003} IMF, solar metallicities, \citet{Madau_1995} inter galactic medium model, \citet{Calzetti_2000} attenuation curve, and parametric SFH (delayed-tau model; $SFR(t) \propto te^{-t/\tau}$). We apply the Markov Chain Monte Carlo using \texttt{emcee} \citep{2013PASP..125..306F} to infer the best-fitting parameters and their associated errors. For dust optical depth and redshift, we use the top-hat prior between $0.0<\tau_V<6.0$ and $0.0<z<5.0$. We utilize the log-uniform prior to the input parameters for $\tau$, age, and stellar mass. To be specific, the parameters are allowed to vary between $0.001<\tau\,\mathrm{[Gyr]}<10.0$, $0.1\times t_\mathrm{univ}<age\,\mathrm{[Gyr]}<t_\mathrm{univ}$, where $t_\mathrm{univ}$ is the age of the Universe at fitted redshift, and $10^{9}<M_*\,[M_\odot]<10^{12}$. 

The right panel of each spectrum in Figure \ref{fig:Specgroup} shows the probability distribution of redshift from SED fitting using \texttt{prospector} (the red lines). Table \ref{tab:specphot} shows the spectrophotometric redshift ($z_\mathrm{specphot}$) from \texttt{prospector} fitting. SFG1, 3, and 4 show the consistent $z_\mathrm{specphot}$ within $1\sigma$ of the QG at $z_\mathrm{spec}=4.53$. SFG2 have slightly different redshift, but we cannot entirely rule out the solution that the galaxy is at the same redshift of the QG due to the low S/N of its spectrum. With 3--4 SFGs located at the same redshift as the central QG and are all within $23''$ (151\,pkpc), we conclude that they form a group environment, making it the highest redshift group with the QG at the center. 

\begin{deluxetable*}{l|ccccccc}
\tablecaption{Spectrophotometric redshift of group members and physical properties inferred from SED fitting. \label{tab:specphot}}
\tablehead{
\colhead{Name} & \colhead{ID\tablenotemark{a}} & \colhead{R.A.} & \colhead{Decl.} & \colhead{$z_\mathrm{specphot}$} & \colhead{$z_\mathrm{spec}$\tablenotemark{b}} &  \colhead{$\theta_{\mathrm{sep}}\,\mathrm{[kpc]}$\tablenotemark{c}} & \colhead{$\log{(M_*/M_\odot)}$\tablenotemark{d}} 
}
\colnumbers
\startdata
    {SFG1} & {1048346} & $\mathrm{10^h02^m27^s.099}$ & $\mathrm{+02^\circ 24' 41''.607}$ & $4.53^{+0.01}_{-0.02}$ & -- & 12.8 & $10.26\pm 0.06$\\ 
    {SFG2} & {1046859} & $\mathrm{10^h02^m26^s.337}$ & $\mathrm{+02^\circ 24' 38''.491}$ & $4.64\pm 0.02$ & -- & 75.0 & $10.06^{+0.11}_{-0.07}$\\ 
    {SFG3} & {1048201} & $\mathrm{10^h02^m27^s.838}$ & $\mathrm{+02^\circ 24' 42''.755}$ & $4.55\pm 0.02$ & -- & 76.4 & $9.74\pm 0.11$\\ 
    {SFG4} & {1041229} & $\mathrm{10^h02^m26^s.318}$ & $\mathrm{+02^\circ 24' 19''.907}$ & $4.56^{+0.09}_{-0.08}$ & -- & 150.8 & $9.89\pm 0.08$\\ 
    \hline
    {QG} & {1047519} & $\mathrm{10^h02^m27^s.091}$ & $\mathrm{+02^\circ 24' 39''.672}$ & -- & $4.5313\pm 0.0005$ & -- & $10.71\pm 0.04$\\ 
\enddata
\tablenotetext{a}{ID number is from COSMOS2020 \textsc{Classic} catalog.}
\tablenotetext{b}{The spectroscopic redshift of the QG is from \citet{Kakimoto2024}.}
\tablenotetext{c}{Projected physical distance between COSMOS-1047519 and SFGs.}
\tablenotetext{d}{Stellar mass is from fitting both spectrum and photometry using \texttt{prospector}.}
\end{deluxetable*}

\subsection{Discussion: Unique environment of the QG} \label{subsec:groupdis}
To further characterize the group, we compare the galaxy stellar mass function with the field value in COSMOS. From \citet{Weaver_2022complete}, the average number density of massive galaxies with $\log{(M_*/M_\odot)}>10^{9.5}$ in COSMOS at $4.5<z<5.5$ is 100 times smaller than that of this group (the galaxy number density is $4.9\times 10^{-2}\, \mathrm{cMpc^{-3}}$, if we define the group volume as a region spanning $151 \times 151\,\mathrm{pkpc}$ in projected plane and $\Delta z \sim 0.03$ along the line of sight). As we discuss in Section \ref{subsec:discgroup}, the final panel of Figure \ref{fig:groups} also indicates that the group has the highest overdensity significance, supporting the group nature of the system. While recent JWST studies also find massive QGs surrounded primarily by lower-mass, emission-line dominated galaxies \citep[e.g.,][]{Suess2023, Carnall2024, Graaff2024, Stawinski2025,Jespersen2025}, our target resides in a notably richer small-scale overdensity characterized by massive members. 

The central QG has a close companion, as do some other known QGs. For example, ``Jekyll \& Hyde'' \citep{Glazebrook2017, Schreiber_2018} is a pair of a QG and an SFG at $z=3.71$ located within a large-scale overdensity \citep{Kakimoto2026}; JWST IFU observations of this system reveal morphological evidence of an ongoing merger \citep{PerezGonzalez_2024}. Similarly, ``Cosmic Vine'' is a large-scale overdensity at $z=3.44$ that hosts a merging pair of QGs \citep[e.g.,][]{Jin2024, Ito_2025, Sillassen2025}, which is expected to eventually collapse to form a single massive galaxy. Interestingly, these galaxy pairs have similar stellar masses and their masses are already very high, suggesting they will undergo (or are undergoing) a major merger. We might be witnessing the evolution of massive (quiescent) galaxies via major mergers in the core of proto-clusters, and these systems finally evolve into massive elliptical galaxies (or brightest cluster galaxies) in the local Universe.

In the local Universe, observations indicate that major mergers trigger AGN activity \citep[e.g.,][]{Brodwin2013,Bickley2023,Ellison2025} and star formation activity, followed by subsequent quenching \citep[e.g.,][]{Gomez-Guijarro2018,Coogan2018,Bickley2022,Ellison2022,Ellison2024}. In the high redshift Universe \citep[$z\gtrsim 5$ when the QG experienced the starburst activity;][]{Kakimoto2024}, bright SFGs are often identified in overdense regions and they often exhibit signs of mergers as revealed by JWST \citep[e.g.,][]{Hashimoto2023, Arribas2024, Hu2025}. The QG and SFG members may be descendants of these galaxies. Our discovery supports that galaxy mergers and/or interactions have a crucial role for evolution of QGs as inferred in one of the rapid quenching modes in \citet{Whitaker2026} at this early epoch.

Our discussion here, however, is based on a small and heterogeneous sample of high redshift QGs. In the following Sections, we are going to statistically address the role of environment for quenching using a large homogeneous sample of galaxies from the COSMOS2020 catalog.

\section{Sample selection and definition of the environment} \label{sec:obs}
To extend this study (specific target) for a large sample of galaxies and wide range of cosmic timescale, we use one of the largest survey regions called COSMOS (Section \ref{subsec:target}). We utilize the multi-wavelength photometric catalog from COSMOS2020 to characterize galaxy environments across redshifts and to put the groups discussed in Section \ref{sec:453} in a broader and more statistical context of environmental effects on galaxy evolution.

\subsection{Sample selection} \label{subsec:follow}
As summarized in \citet{Ito_2022}, we applied our photo-$z$ code \citep[\texttt{MIZUKI};][]{Tanaka_2015} to the COSMOS2020 \textsc{Classic} catalog. The code adopts the \citet{BC03} models, assuming exponentially declining SFHs ($SFR(t) \propto e^{-t/\tau}$), solar metallicities, and the \citet{Calzetti_2000} dust attenuation curve. From the comparisons between photometric and spectroscopic redshifts ($\Delta z = |z_\mathrm{phot}-z_\mathrm{spec}|$), our photometric redshifts are confirmed to be accurate for massive QGs at $2<z<4$ in the COSMOS field ($\sigma_\mathrm{NMAD}[\Delta z/(1+z_\mathrm{spec})]=0.03, \mathrm{Median}(\Delta z)=0.011,\eta = 0.0\%$ where $\eta$ is outlier rate of galaxies with $\Delta z/(1+z_\mathrm{spec})>0.15$; \citealp{Ito_2022}, and references therein). Furthermore, the accuracy of the photometric redshifts are also high for all galaxies at $1<z<5$ in the COSMOS field ($\sigma_\mathrm{NMAD}[\Delta z/(1+z_\mathrm{spec})]=0.024, \mathrm{Median}(\Delta z)=0.008, \eta = 6.5\%$) using the COSMOS galaxy $z_\mathrm{spec}$ catalog \citep{Khostovan2025}.

\begin{figure}
    \plotone{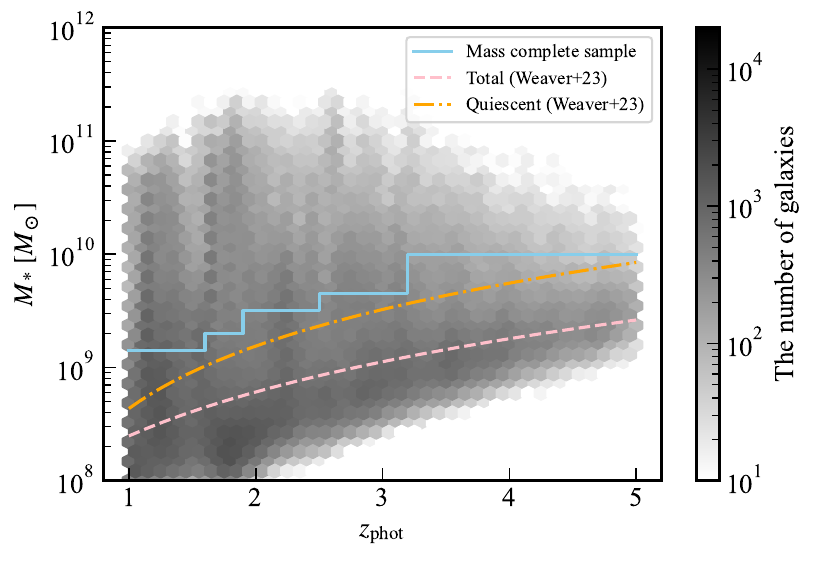}
    \caption{Stellar mass 70\% completeness in COSMOS2020 (all galaxies: the pink line, QGs: the orange line; \citealp{Weaver_2022complete}). The sky blue line shows the selected completeness limit for each redshift bin (Section \ref{subsec:follow}). The dark color map shows the galaxy numbers at each redshift and stellar mass.}
    \label{fig:complete}
\end{figure}

Based on this highly accurate photo-$z$ catalog, we applied the following cuts to select galaxies for our main analysis; (1) $1<z_\mathrm{phot}<5$, (2) the target is sufficiently bright for us to trace its continuum (satisfy 70\% mass completeness in the COSMOS2020; \citealp{Weaver_2022complete}; the sky blue line in Figure \ref{fig:complete}), (3) $M_\mathrm{*} < 10^{11.5}\,M_\odot$ to exclude galaxies with potentially failed photo-$z$ estimates, and (4) reduced-$\chi^2$ value is smaller than 5. We note that the results are not changed if we take (3) criteria with $M_\mathrm{*} < 10^{12}\,M_\odot$.
For selecting QGs, we select galaxies whose 1 sigma upper limit of SFR is one dex below the star-forming main sequence (SFMS) from recent research \citep{Popesso2023} to acquire the conservative results. We also conduct the same analysis with QGs selected with $UVJ$ \citep{Williams_2009} and $NUVrJ$ \citep{Ilbert2013} color and confirmed that our results remain consistent (see Appendix \ref{sec:ApeA} in detail).

\begin{deluxetable*}{l|cc}
\tablecaption{The number of massive galaxies and QGs at each redshift range. \label{tab:numgal}}
\tablehead{
\colhead{$z_\mathrm{phot}$} & \colhead{$M_*>10^{10}\,M_\odot$} & \colhead{QG} 
}
\colnumbers
\startdata
    $1.1<z<1.6$ & 11109 & 2674 \\ 
    $1.6<z<1.9$ & 9952 & 1759 \\ 
    $1.9<z<2.5$ & 10422 & 1012 \\ 
    $2.5<z<3.2$ & 10119 & 681 \\ 
    $3.2<z<5.0$ & 9307 & 157 \\ 
\enddata

\end{deluxetable*}

\subsection{Definition of the environment} \label{sec:redshift}
Galaxy surface overdensity value is defined by the following equation:
\begin{equation} \label{eq:delta}
    \delta = \frac{n-\bar{n}}{\bar{n}},
\end{equation}
where $n$ is the galaxy surface density, and $\bar{n}$ is the averaged surface density in the entire COSMOS field at each redshift slice (see the definition of the redshift slice below). In this study, the galaxy surface density is computed with the $k$-th nearest neighbor method ($k$-th NN). This method measure the projected distance from a target galaxy to its $k$-th nearest neighbor. The $k$-th galaxy can be included in the galaxy number count within a circle whose radius is this distance. NN is indicated to be the most useful approach for tracing the local environment of galaxies \citep{Muldrew2012, Chiang2013}. In addition, we can easily define a different scale of environment with a different $k$. The surface density of $k$-th NN ($n_k$) can be expressed by the following equation:
\begin{equation} \label{eq:NN}
    n_k = \frac{k}{\pi r_k^2}.
\end{equation}
Here, $r_k$ is the projected distance to $k$-th nearest galaxy. 

Statistical uncertainties of our photometric redshifts vary with redshift (they are larger at larger redshifts). To account for this variation, we utilized the following formula to set the range of photometric redshifts of galaxies to be used in the mapping process at each redshift slice. 
\begin{equation} \label{eq:dz}
    \Delta z = 0.036 \times (1+z)
\end{equation}
The value of 0.036 is obtained from 1.5 times the scatter between $z_\mathrm{phot}$ and $z_\mathrm{spec}$ (Section \ref{subsec:follow}). Since the accuracy of photometric redshifts degrades toward higher redshifts, the width of the redshift slice used for the density estimation is designed to be redshift-dependent. We compute the mean surface density of galaxies over the entire COSMOS field in each redshift bin (Equation \ref{eq:dz}) with a spacing of $z = 0.05$. We then draw the overdensity maps of galaxies using Equation \ref{eq:delta}.

To make a fair definition of environment over the wide redshift range considered here, we use only massive galaxies with $M_*>10^{10}\,M_\odot$ to compute the galaxy surface density. Lower mass galaxies are included in our analysis below, but the environment is defined using this mass-selected subsample of galaxies. Table \ref{tab:numgal} shows the number of massive galaxies at each redshift range. The range of each redshift is chosen to include comparable numbers of massive galaxies to compare properly. We note that if we use an evolving mass cut at a constant cumulative number density of galaxies \citep{Behroozi2013} using the same catalog to account for the mass evolution of galaxies at $1<z<5$, the following results do not significantly change but the physical scale of overdensity becomes larger at lower redshift due to small number of massive galaxies in each redshift slice.

\section{Environmental dependence of quiescent galaxy fraction} \label{sec:SEDfit}
In this study, we utilize the 3rd NN ($k=3$). This is because we would like to explore environments at small scales (fewer companion galaxies) such as group as found in Section \ref{sec:453}.

\subsection{Environmental quenching} \label{subsec:SED}
\begin{figure}
    \plotone{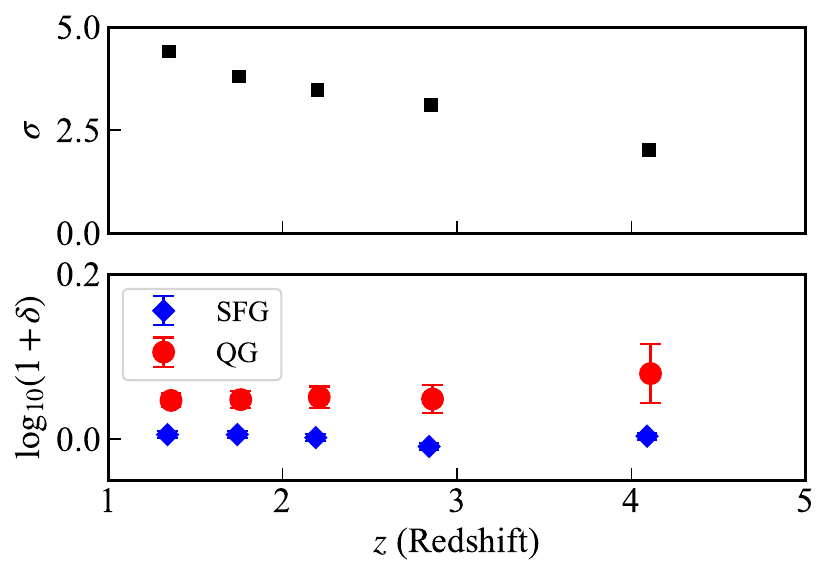}
    \caption{Redshift evolution of overdensity values for massive QGs and SFGs ($M_*>10^{10}\,M_\odot$). Top Panel: The significance of overdensity value for QGs against the average overdensity of SFGs according to the bins defined in Table \ref{tab:numgal}. Here, $\sigma = [\delta(\mathrm{QG})-\delta(\mathrm{SFG})]/\sqrt{\sigma_\delta(\mathrm{QG})^2+\sigma_\delta(\mathrm{SFG})^2}$ where $\sigma_\delta$ is a standard deviation of $\delta$. Bottom Panel: The average and the standard deviation of overdensity values at $1<z<5$.}
    \label{fig:zevo}
\end{figure}

Figure \ref{fig:zevo} shows the average and standard deviation of the galaxy number density around SFGs and QGs at $1<z<5$ (where the average number density in COSMOS is $\delta = 0$). We find that the average overdensity of QGs is higher than that of SFGs by 3--5$\sigma$ at $z\lesssim 3$, suggesting that QGs are preferentially located in overdense regions. The significance is decreased at $z>3.2\ (\sim 2\sigma)$, but the trend is still consistent with the surrounding environment of spectroscopically confirmed QGs at $z > 4$ being denser than average \citep[e.g.,][]{Carnall2024,Graaff2024, McConachie2025}. On the other hand, Figure \ref{fig:zevo} also demonstrates that QGs in galaxy groups ($\log{(1+\delta)}>1$) such as the one discovered in Section \ref{sec:453} are not common (see Appendix \ref{sec:ApeB} for examples of groups). Even considering the uncertainty of photometric redshifts and the small sample size of QGs, the galaxy group in Section \ref{sec:453} is likely among the highest density regions discovered to date. In addition, this is the highest-redshift galaxy group containing a QG in the COSMOS field.

\begin{figure*}[tb]
    \centering
    \begin{minipage}{0.49\textwidth}
        \centering
        \includegraphics[width=0.95\textwidth]{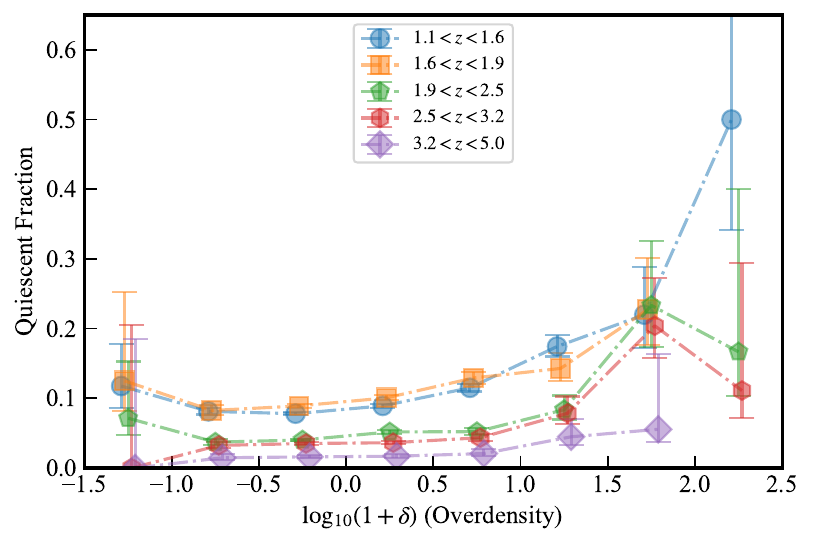}
    \end{minipage}
    \begin{minipage}{0.49\textwidth}
        \centering
        \includegraphics[width=0.95\textwidth]{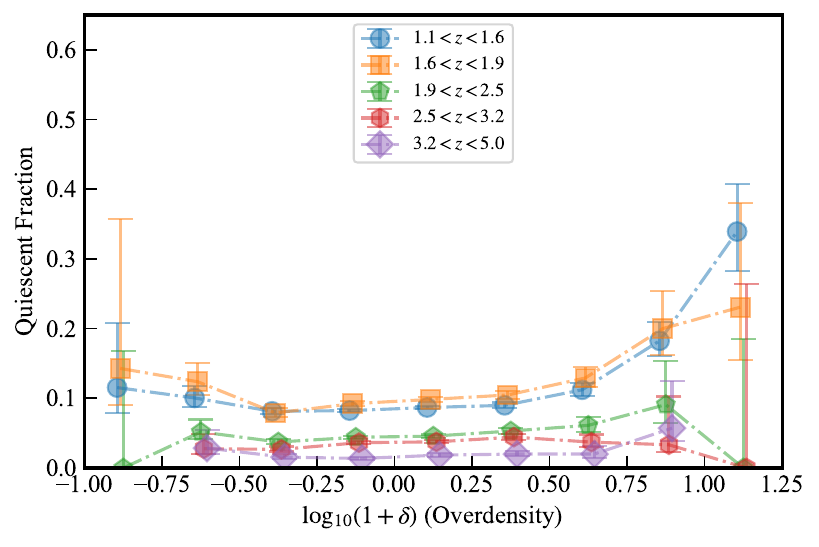}
    \end{minipage}
    \caption{Left Panel: Fraction of QGs in overdensity bins (3rd nearest neighbor method). The uncertainty of the quiescent fraction is computed from (Bayesian) binomial confidence intervals \citep{Cameron2011}. The fraction is shown only if there are more than four galaxies at each density bin. Right Panel: Fraction of QGs in overdensity bins (10th nearest neighbor method).}
    \label{fig:QF3rd}
\end{figure*}

The left panel of Figure \ref{fig:QF3rd} shows the fraction of QGs relative to all galaxies in each bin as a function of overdensity. Different colors show different redshift ranges summarized in Table \ref{tab:numgal}. The fraction of QGs is higher in overdensity region (especially $\log{(1+\delta)}>1$) than in the underdensity field at all redshifts. The overdensity of $\log{(1+\delta)}\sim1$--$1.5$ corresponds to $r_3\sim 200\,\mathrm{pkpc}$, independent of the redshift range, and this scale is the typical virial radius $(R_{200})$ of the proto-cluster core \citep{Chiang_2017}. This result suggests that the formation of QGs is more accelerated in denser environment even in the distant Universe. Thus, the small-scale overdensity environment contributes to galaxy quenching even at high redshift.

In the right panel of Figure \ref{fig:QF3rd}, the 10th nearest neighbor method is applied to trace the proto-cluster scale environment ($r_{10} \sim 500\mathrm{-}900\,\mathrm{pkpc}$). At $1.9<z<5$, we can see that there is no significant correlation between overdensity and quiescent fraction. This indicates that the small-scale (local) environment is more important to change the galaxy population at high redshift. On the other hand, there is a clear enhancement of quiescent fraction between field and overdensity region at $z<1.9$. This is possibly because the large-scale structure develops and clusters become richer at lower redshifts. Within these virialized massive halos, the gas is stably shock-heated to the virial temperature as a consequence of stable shock propagation in halos above a critical mass threshold \citep[e.g.,][]{Keres2005, Dekel2006}. In these hot-gas environments, processes such as ram-pressure stripping efficiently deplete the star-forming gas in infalling low-mass galaxies \citep[e.g.,][]{Peng_2010, Wetzel2013}.

\subsection{Mass quenching} \label{subsec:property}
\begin{figure}
    \plotone{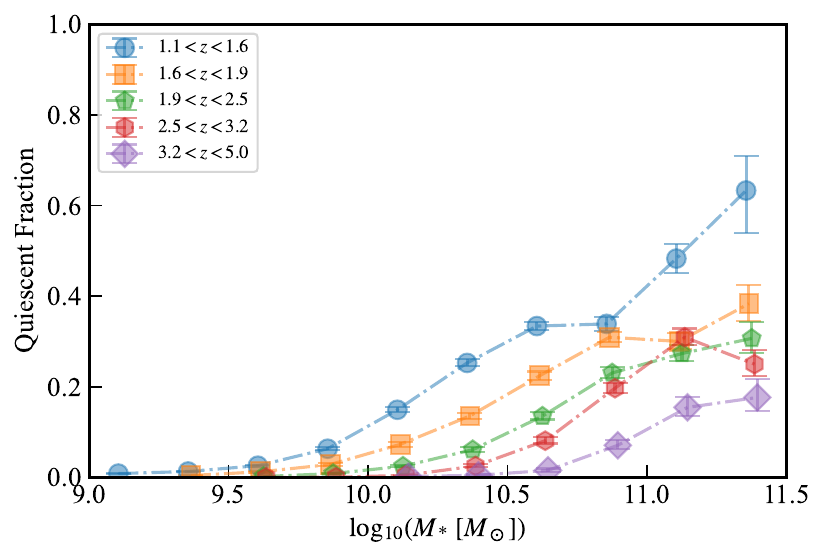} 
    \caption{Fraction of QGs in stellar mass bins. The uncertainty is as in Figure \ref{fig:QF3rd}.}
    \label{fig:massQF}
\end{figure}
While environment is one of the key drivers for quenching, stellar mass of galaxies is another key; most massive galaxies are quenched regardless of environment \citep[e.g.,][]{Peng_2010}. Figure \ref{fig:massQF} shows the fraction of QGs in stellar mass bins at $1<z<5$. At all redshifts, galaxies with larger stellar masses have a higher fraction of QGs. This figure clearly illustrates that mass quenching is also occurring in the early Universe up to $z=5$. In addition, the figure demonstrates that the fraction of low-mass QGs ($M_*\lesssim 10^{10.5}\,M_\odot$) increases at lower redshifts, while the fraction is nearly zero at high redshifts. This change is consistent with the downsizing scenario observed in the SFHs of elliptical galaxies in the local Universe \citep{Gallazzi_2005, Thomas2010, Gallazzi2025}, in which lower mass galaxies are quenched at lower redshifts (in other words, the quenching of massive galaxies occurs at higher redshifts). We will further discuss this trend in the next subsection.

\subsection{Comparison with the QG fraction in the local Universe} \label{subsec:Peng10}
\begin{figure}[tb]
    \plotone{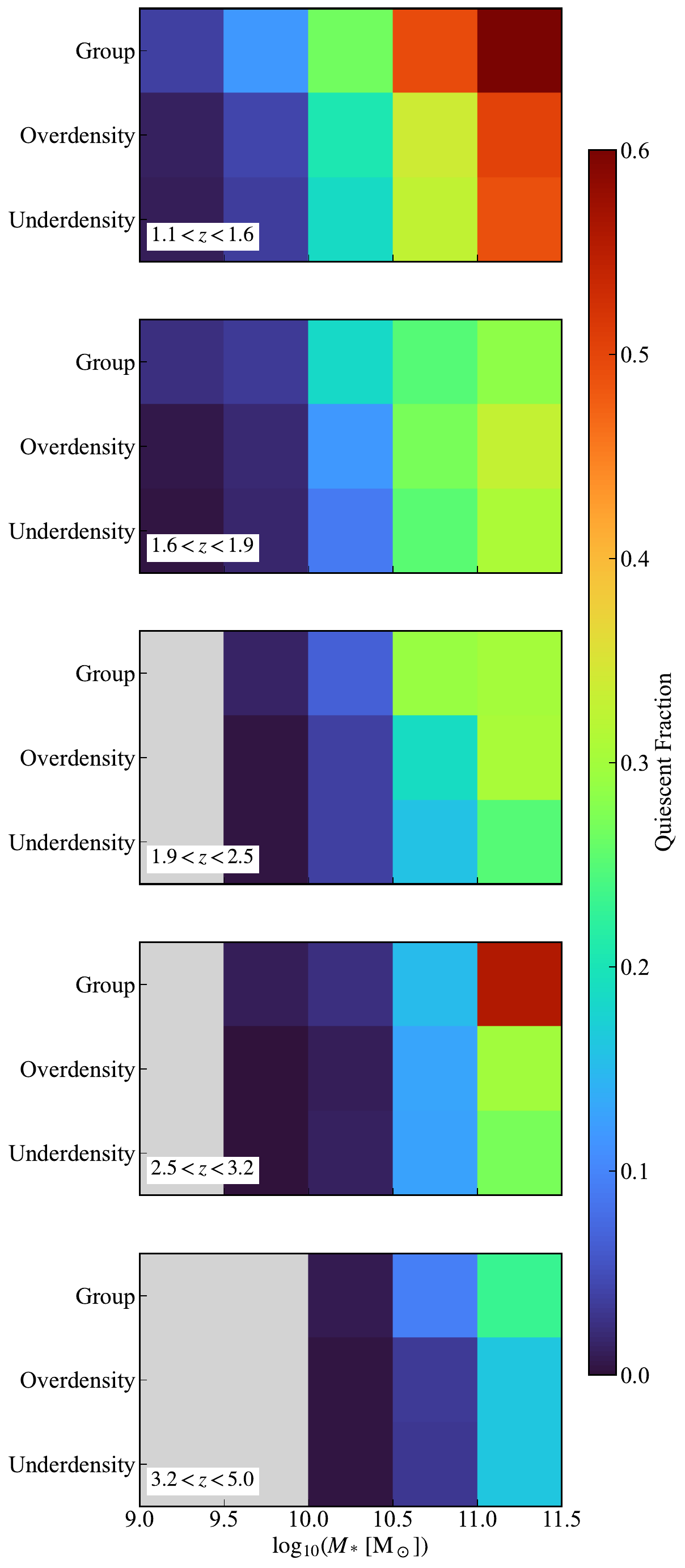}
    \caption{QG fraction as functions of stellar mass and environment at $1<z<5$. Environment is defined by three types (Group: $\log{(1+\delta)}>1$, Overdensity: $0<\log{(1+\delta)}<1$, and Underdensity: $\log{(1+\delta)}<0$). The rainbow color shows the quiescent fraction at each stellar mass bin and density bin. The gray region represents no galaxies due to our sample selection (Section \ref{subsec:follow}).}
    \label{fig:peng}
\end{figure}

Finally, we plot the QG fraction as functions of overdensity and stellar mass in Figure \ref{fig:peng}, extending the work by \citet{Peng_2010} to the higher redshifts. The rainbow color shows the QG fraction at each stellar mass and surrounding galaxy number density. We divide the three types of environment, underdensity ($\log{(1+\delta)}<0$), overdensity ($0<\log{(1+\delta)}<1$), and group ($\log{(1+\delta)}>1$). The last one corresponds to the significant local overdensity region. In order to fairly compare our results with those of \citet{Peng_2010}, we adopt this threshold where the literature shows a significant difference in the red fraction compared to the field. If we vary the group threshold within $1.0\pm 0.2$ \citep[such as $0.8$ used in][]{Darvish2016}, the trends in Figure \ref{fig:peng} do not change significantly.

\begin{figure}[tb]
    \plotone{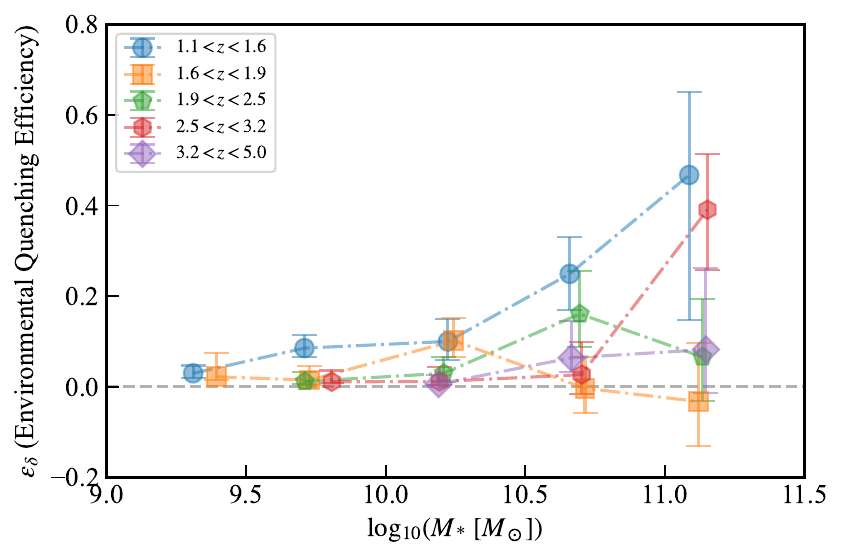}
    \caption{Environmental quenching efficiency of galaxies as a function of stellar mass.}
    \label{fig:Qeff}
\end{figure}

At $1.1<z<2.5$, a clear correlation with the fraction of QGs can be seen in columns and rows with different masses and galaxy densities. Therefore, both mass and environment are the drivers of galaxy quenching in this redshift regime. This is broadly consistent with the \citet{Peng_2010} result, except for low-mass QG fraction (the fraction of low-mass QGs ($M_*<10^{10}\,M_\odot$) in overdensity is not as high as massive QGs; \citealp{Baldry2006}). This is also seen in the enhancement of the environmental quenching efficiency defined in \citet{Peng_2010} with more than 95\% significance level in Figure \ref{fig:Qeff}. We utilize the QG fraction ($f_\mathrm{Q}$) in underdensity ($\delta_0$) and group region to constrain the efficiency ($\epsilon_\delta$) defined by following equation,
\begin{equation} \label{eq:EQF}
    \epsilon_\delta =\frac{f_\mathrm{Q}(\delta, m, z)-f_\mathrm{Q}(\delta_0, m, z)}{f_\mathrm{SF}(\delta_0, m, z)},
\end{equation}
where $f_\mathrm{SF}(\delta_0, m, z) = 1-f_\mathrm{Q}(\delta_0, m, z)$.

However, galaxies with $M_*<10^{10.5}\,M_\odot$ are mainly SFGs and the most massive galaxies only show the apparent excess of QG fraction in the overdensity region at $z>2.5$. This is also suggested in the environmental quenching efficiency, which indicates that more massive galaxies have stronger dependence with the galaxy surrounding environment.

Interestingly, Figure \ref{fig:peng} also represents the downsizing trend at $1<z<5$ that we briefly mentioned in the last subsection. In particular, the quiescent fraction at $3.2<z<5.0$ and that of $M_*<10^{11}\,M_\odot$ galaxies at $1.9<z<2.5$ are very similar in all environments. The similar QG fractions can also be found at even a lower redshift and lower mass range ($M_*<10^{10.5}\,M_\odot$ and $1.1<z<1.6$). Figure \ref{fig:peng} nicely illustrates that the lower mass galaxies are quenched at a later epoch.

The down-sizing trend can also be appreciated in Figure \ref{fig:Qeff}, which shows the similar $\epsilon_\delta$ in $M_*>10^{11}\,M_\odot$ galaxies at $3.2<z<5.0$, $10^{10.5}\,M_\odot<M_*<10^{11}\,M_\odot$ galaxies at $1.9<z<2.5$, and $10^{10}\,M_\odot<M_*<10^{10.5}\,M_\odot$ galaxies at $1.1<z<1.6$. These results again suggest the mass (and redshift) dependent environmental effects. Specifically, we expect a downsizing trend wherein accelerated formation of QGs occurs at high redshifts through processes such as mergers, while the suppression of star formation at lower redshifts is driven by environmental quenching that becomes effective in lower mass galaxies. Consequently, this leads to the time evolution of the quiescent fraction as a function of stellar mass, as indicated in Figure \ref{fig:massQF}.

To be more quantitative about the excess of QG fraction for massive galaxies and galaxies in the overdensity regions regardless of the definitions of stellar mass or density bins, we compare the correlation function between QG fraction and stellar mass or number density using logistic regression analysis. The probability that a galaxy is quiescent is denoted by $p(Y=1)$ (where $Y$ is a response variable), and explanatory variables are defined by $x_1 = \log{(M_*/M_\odot)}$ and $x_2 = \log{(1+\delta)}$. If we assume the partial regression coefficient of each explanatory variable as $a_0, a_1$, and $a_2$, respectively, the logistic regression model is expressed by the following equation.
\begin{equation} \label{eq:logistic}
    \log{\left(\frac{p(Y=1)}{1-p(Y=1)}\right)} = a_0+a_1x_1+a_2x_2.
\end{equation}
In order to estimate the partial regression coefficients, which indicates the importance of mass quenching ($a_1$) and environmental quenching ($a_2$), we utilize the maximum likelihood method:
\begin{equation} \label{eq:likelihood}
    L = \prod_i P_i^{Y_i} (1-P_i)^{(1-Y_i)},
\end{equation}
where $L$ is the likelihood, $i$ represents over all galaxies in each redshift bin, and $P_i = p(Y_i=1)$.

\begin{figure}
    \plotone{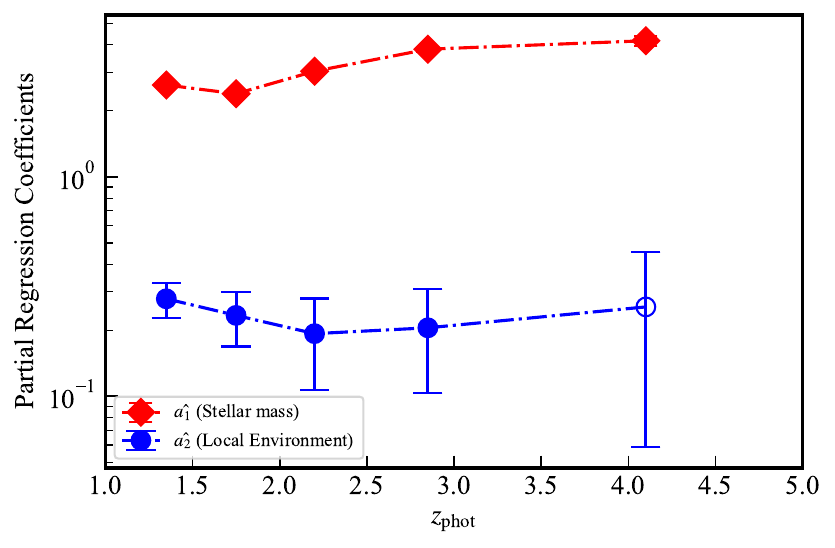} 
    \caption{Partial regression coefficients from logistic regression analysis for mass quenching ($\hat{a_1}$) and environmental quenching ($\hat{a_2}$) as a function of redshift (galaxy number density is from 3rd NN method). Both of stellar mass and environment have positive coefficients, which indicates that QGs tend to be more massive and located in denser environment. Filled circle indicates that the coefficients are positive at more than 95\% significance level, while the open circle means less than it.}
    \label{fig:Log}
\end{figure}

Figure \ref{fig:Log} compares the maximum likelihood estimates of the partial regression coefficients ($\hat{a_1}, \hat{a_2}$) in the analysis. The results show positive correlations for all parameters even at high redshift. As shown in Figures \ref{fig:QF3rd} and \ref{fig:massQF}, the correlation with stellar mass is stronger than environment. However, even considering the mass correlation by applying the multivariable logistic regression, correlation of environment and QG fractions are confirmed at $1<z<3.2$ with more than 95\% significance level. In spite of no significant correlation of number density and QG fraction at $z>3.2$ due to the small number of QGs, the partial regression coefficients is still consistent with the lower redshift values. These results suggest that the environmental quenching also occurs at least up to $z\sim 3$ and possibly beyond. In the following, we discuss the possible interpretations of these results together with results from the literature.

\section{Discussion}\label{sec:group}
\subsection{Galaxy group at $z=4.53$}\label{subsec:discgroup}
Firstly, we discuss how the group shown in Section \ref{sec:453} fit the context of the evolution of QGs. Figure \ref{fig:groups} indicates the group (the final panel) represents the densest region of galaxies containing a QG located at the highest redshift within the COSMOS field. Therefore, this group is not common; rather, it is an example of accelerated evolution of massive galaxies by overabundance of galaxies inside the small-scale overdensity. Figures \ref{fig:QF3rd} and \ref{fig:peng} commonly show the correlation that quenching efficiency becomes higher in small-scale overdensity (such as galaxy group), and thus the spectroscopically confirmed group in Section \ref{sec:453} is an example of such extreme environment at high redshifts.

In addition, the existence of a companion galaxy around the QG might indicate that major mergers and/or interactions are crucial to form such a massive QG at high redshift. Recall that there are similar cases with close companions such as Jekyll \& Hyde and QG pair in Cosmic Vine \citep{Jekyll2018,Jin2024}. Deep and high resolution imaging data (such as from JWST) is required to confirm the evidence of mergers or interactions of the QG at $z=4.53$. 

We note that this galaxy is not classified as a QG when relying solely on constraints from photometric information (it has a suppressed SFR but does not fully satisfy our QG criteria), suggesting that there remains a possibility that potential QGs at high redshifts are being missed in current photometric selections. Moreover, because the QG solution was determined more robustly by incorporating constraints from the rest-frame optical spectrum \citep{Kakimoto2024}, obtaining such spectral constraints plays a crucial role at high redshifts.

\subsection{Comparison with the previous research} \label{sec:disc}
At $z\sim 1$, the environmental dependence of QGs has been clearly observed, confirming an excess of quiescent fraction in galaxy cluster regions relative to the field and the high efficiency of environmental quenching \citep[e.g.,][]{Patel2009,Scoville2013,Kawinwanichakij2017,Burg2020,Werner2022}. Our results confirm the same correlation as shown in Figure \ref{fig:peng} and \ref{fig:Log}.

At cosmic noon, previous research has confirmed a correlation between the QG fraction and overdensity (or the redshift evolution of environmental quenching efficiency) over a wide redshift range. Some studies indicate that the QG fraction is higher in overdensity regions at redshifts up to $z=1$ \citep{Darvish2016} and $z=2$ \citep{Shi2024}. Environmental quenching efficiency (equation \ref{eq:EQF}) is also significant at up to $z=1$ \citep{Shi2024}, $z=2$ \citep{Taamoli2024}, and $z\sim 3$ \citep{Darvish2016, Chartab2020}. At lower redshifts ($z<2$), previous research shows consistent results—where overdensity regions typically host more QGs than the field—but the cosmic epoch at which environmental quenching is no longer observed is still under discussion. For instance, \citet{Singh2024} defined the galaxy surface number density using LAEs selected with the narrow-band imaging technique at $z=3.1$. Based on the lack of correlation between the QG fraction and overdensity values, they concluded that the SFR is independent of the environment. However, recent JWST observations have begun to confirm the presence of massive QGs located within significant overdensities \citep[e.g.,][]{McConachie2025}.

Our results show the significant correlation between QG fraction and overdensity up to $z=3$ and suggest the possible correlation even at $z>3$. The most important point of our study is the definition of the scale of galaxy surrounding environment. Although the correlation between the overdensity and quiescent fraction is clearly seen in the entire redshift range explored with the 3rd NN method which traces the local overdensity (the left panel of Figure \ref{fig:QF3rd}), the correlation disappears at $z>1.9$ in the 10th NN method, which traces the larger scale environment (the right panel of Figure \ref{fig:QF3rd}). This redshift range of disappearing the QG fraction-density relation is consistent with \citet{Shi2024}, and the environment proved by previous studies is primarily a large-scale environment. Thus, this study highlights the important role the small-scale environment ($\sim 100\,\mathrm{pkpc}$) plays in high-redshift galaxy quenching and/or the evolution of massive QGs. \citet{Pan2025} suggested the no evidence of environmental quenching in the proto-cluster at $z=2.6$, which is consistent with the small $\epsilon_\delta$ in lower mass galaxies in the left panel of Figure \ref{fig:Qeff} although we cannot trace the environment of low-mass ($M_*<10^{9.5}\,M_\odot$) galaxies. We only show the significant level of $\epsilon_\delta$ in the most massive galaxies but \citet{Pan2025} are not able to include such galaxies maybe due to the low statistics.

In \citet{Edward2024}, the efficiency of environmental quenching is summarized at $0<z<2$ based on previous discoveries of galaxy clusters and other regions. They suggest that the efficiency of environmental quenching decreases with increasing look-back time, which is generally consistent with the trend of the partial regression coefficients in Figure \ref{fig:Log}. Some case studies show that environmental quenching efficiency persists up to $z\sim 3$, which agrees with our findings in Figure \ref{fig:Qeff}. However, the positive correlation between environmental quenching efficiency and stellar mass \citep{Darvish2016,Chartab2020,Taamoli2024} suggests the presence of mass quenching inside overdensity and the role of mass segregation \citep[e.g.,][]{Oemler1974, Kauffmann2004, Baldry2006} in the evolution of QGs. It is important to confirm more proto-cluster regions identified by QGs in the distant Universe \citep[e.g.,][]{McConachie_2022,Ito_2023,Tanaka2024} in order to interpret this environmental dependence.

\subsection{Mechanism of galaxy quenching in the overdensity} \label{sec:phys}
From these results, we finally attempt to address the formation scenario of massive QGs at high redshift. Galaxy mergers and interactions occur more frequently in small-scale overdensity at high redshift, which induces environmental dependence of the quiescent fraction (Figure \ref{fig:QF3rd}) due to merger-induced starburst and merger-induced AGN \citep[e.g.,][]{Hopkins2008,Lotz2013,Hine2016,Shibuya2025}. These mergers and interactions are also required to form the most massive galaxies at high redshifts in cosmological simulations \citep[e.g.,][]{Qin2017,Xie2024,McConachie2025}. Then, the AGN kinetic mode feedback and/or depletion of surrounding gas due to their starburst may contribute to stop their star formation activity. The dependence of environmental quenching efficiency on stellar mass indicates that environmental quenching mechanisms for low-mass QGs discussed in the low redshift \citep[such as strangulation and ram-pressure stripping; e.g.,][]{Larson1980,Balogh2000,Gunn1972,Boselli2016} are not dominant, but we need more low-mass QGs to identify the physical mechanisms of environmental quenching as a function of stellar mass at high redshift. 
This quenching and formation scenario of QGs inside the overdensity is similar to the evolution scenario of proto-cluster core discussed in previous research \citep{Chiang_2017, Shimakawa_2018, Hewitt2025}. 

At lower redshift, the mass quenching process occurs and lower-mass galaxies tend to be quenched due to the environmental effects in (proto-)cluster, which increases the QG fraction at all masses. This downsizing trend of $\epsilon_\delta$ and quiescent fraction as a function of stellar mass (Figures \ref{fig:QF3rd}, \ref{fig:massQF}, and \ref{fig:Qeff}) indicates the importance of different environmental mechanisms at low redshift compared to high redshift. Within already virialized clusters, processes such as gas stripping and interaction with the hot-gas environment efficiently deplete the star-forming gas in low-mass galaxies. These varying environmental effects over cosmic time provide a plausible explanation for the observed downsizing of quiescent galaxies.

\section{Conclusion} \label{sec:Concl}
In this study, we spectroscopically confirm the galaxy group around a QG at $z=4.53$ discovered in \citet{Kakimoto2024}. The follow-up observation of the Lyman break feature using Subaru/FOCAS allow us to constrain the same spectrophotometric redshifts of three SFGs with the QG. The QG is confirmed to be located in the small-scale overdensity region (at least three SFGs are located within 150\,pkpc from the QG).

In addition, we explore the dependence of the abundance of QGs in the COSMOS field on stellar mass and the density of surrounding galaxies (environment) to confirm whether the group environment is common for QGs or not. At $1<z<3.2$, even when considering the mass dependence of the QG fraction, which is clearly significant (60--20$\sigma$), the environment dependence is still confirmed (4--2$\sigma$). This suggests that “environmental quenching” may also be at work in the Universe beyond cosmic noon at $z\sim 3$. On the other hand, at $3<z<5$, although QGs existing in galaxy group environments such as \citet{Kakimoto2024} are confirmed, the environmental dependence of the QG fraction is not very strong. However, comparisons of the partial regression coefficients and the average galaxy number density in regions where QGs exist suggest that environmental dependence may still exist, necessitating further confirmation of QGs at high redshift.

Our results indicate that small-scale environment shows a stronger correlation with the QG fraction especially at $z>2$ from comparison with previous research. At these redshifts, the group-like environment likely enhances galaxy mergers and interactions, which induce starbursts and/or AGN activity in massive galaxies, eventually leading to rapid quenching. Thus, the presence of massive QGs within the overdensity suggests that such environmental effects play a crucial role in the formation of QGs at high redshifts. On the other hand, at $z<2$, we confirm the similar correlation of the QG fraction in overdensity bins in larger-scale and increased fraction of low-mass QGs, suggesting that environmental quenching such as gas stripping which affect especially the low-mass galaxies may crucial. This time evolution of environmental quenching may trigger the downsizing trend of QG fractions at $1<z<5$.

\begin{acknowledgments}
    We thank the anonymous referee for constructive comments. TK acknowledges support from JSPS grant 25KJ1331. This work was supported by the Data‐Scientist‐Type Researcher Training Project of The Graduate University for Advanced Studies, SOKENDAI. FV and KI acknowledge support from the Independent Research Fund Denmark (DFF) under grant 3120-00043B. The Cosmic Dawn Center (DAWN) is funded by the Danish National Research Foundation (DNRF140). K.M. acknowledges the Waseda University Grant for Special Research Projects (Project number: 2025C-484). This work was partially supported by Overseas Travel Fund for Students (2024, 2025) of Astronomical Science Program, The Graduate University for Advanced Studies, SOKENDAI. 
    This research is based in part on data collected at the Subaru Telescope, which is operated by the National Astronomical Observatory of Japan. We are honored and grateful for the opportunity of observing the Universe from Maunakea, which has the cultural, historical and natural significance in Hawaii.
    Based on observations collected at the European Southern Observatory under ESO programme ID 179.A-2005 and on data products produced by CALET and the Cambridge Astronomy Survey Unit on behalf of the UltraVISTA consortium.
\end{acknowledgments}

%

\vspace{5mm}
\facilities{Subaru (FOCAS)}


\software{Astropy \citep{2013A&A...558A..33A,2018AJ....156..123A},
          \texttt{emcee} \citep{2013PASP..125..306F},
          IRAF \citep{Tody1986, Tody1993},
          Matplotlib \citep{Hunter:2007},
          \texttt{MIZUKI} \citep{Tanaka_2015},
          numpy \citep{harris2020array},
          \texttt{prospector} \citep{Johnson_2021},
          Python-fsps \citep{Conroy_2009,Conroy_2010}
          }

\appendix

\section{QG selection method} \label{sec:ApeA}
\citet{Forrest2024} show the difference of selection criteria for QGs make some results change such as environmental dependence. There are several criteria for selecting QGs at high redshift such as $NUVrJ$ color \citep[e.g.,][]{Ilbert2013,Darvish2016, Taamoli2024}, $UVJ$ color \citep[e.g.,][]{Williams_2009,McConachie2024}, and $sSFR$ selection \citep[e.g.,][]{carnall_2023, Singh2024}. The QGs and SFGs selected by these different methods overlap substantially, although subtle differences are observed.

\begin{figure*}[tb]
    \centering
    \begin{minipage}{0.49\textwidth}
        \centering
        \includegraphics[width=0.95\textwidth]{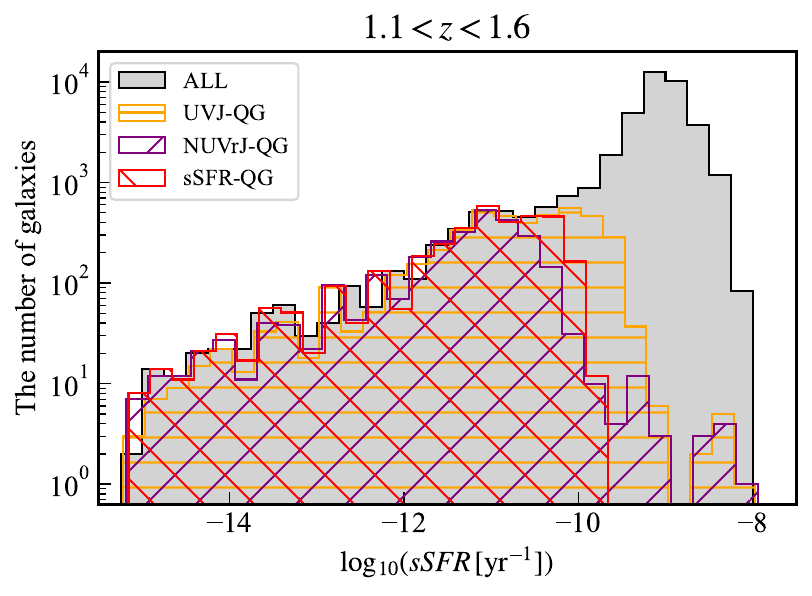}
    \end{minipage}
    \begin{minipage}{0.49\textwidth}
        \centering
        \includegraphics[width=0.95\textwidth]{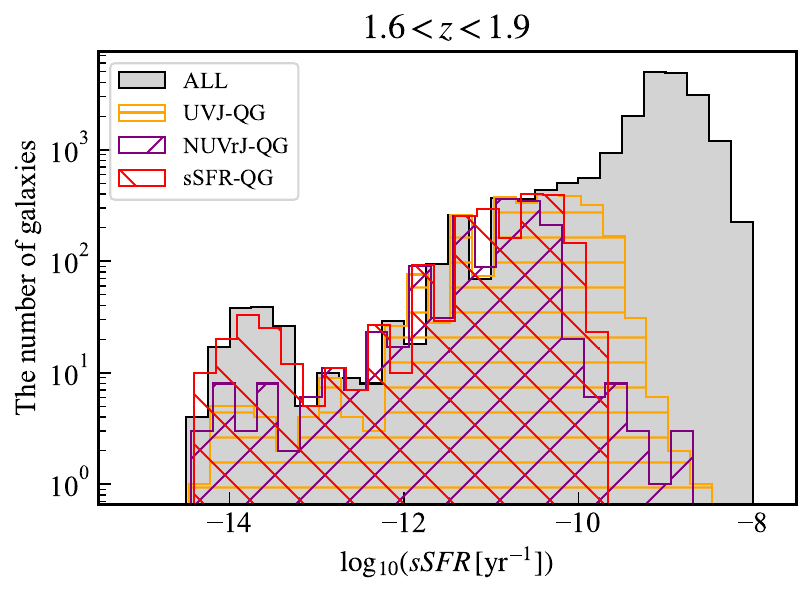}
    \end{minipage}
    \begin{minipage}{0.49\textwidth}
        \centering
        \includegraphics[width=0.95\textwidth]{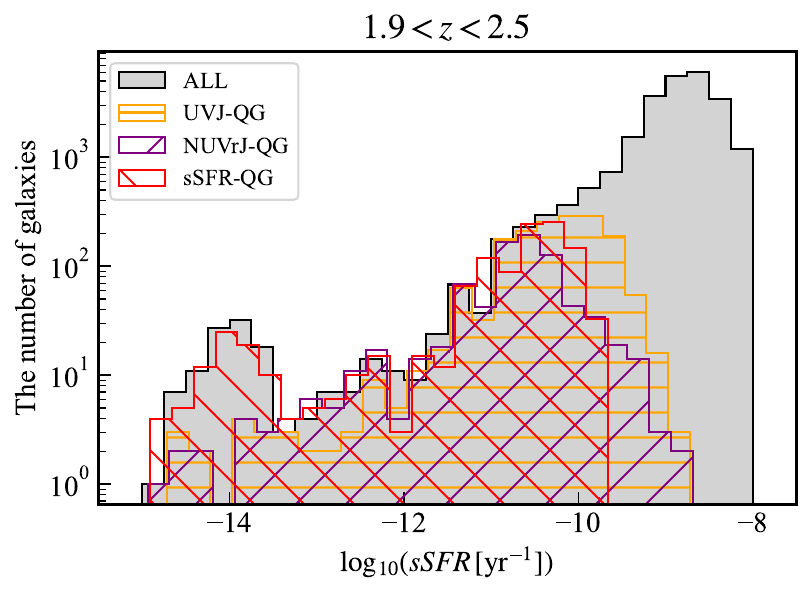}
    \end{minipage}
    \begin{minipage}{0.49\textwidth}
        \centering
        \includegraphics[width=0.95\textwidth]{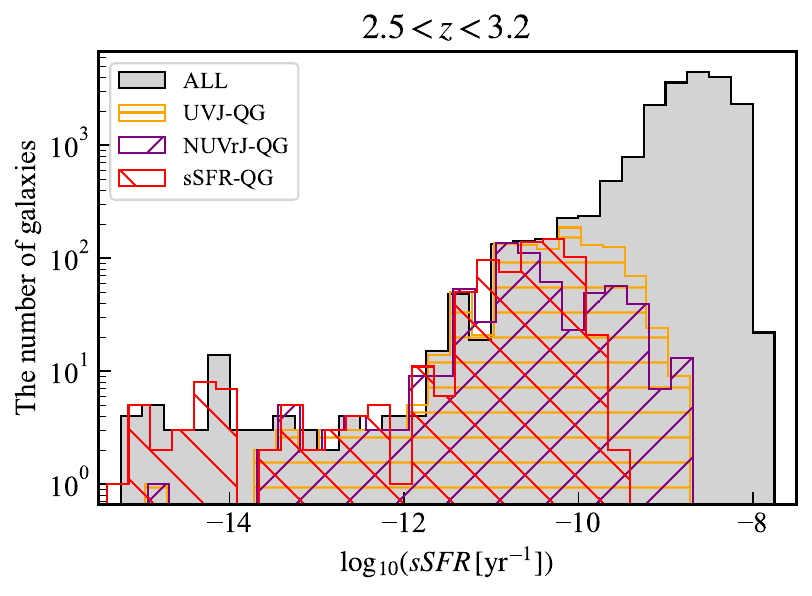}
    \end{minipage}
    \begin{minipage}{0.49\textwidth}
        \centering
        \includegraphics[width=0.95\textwidth]{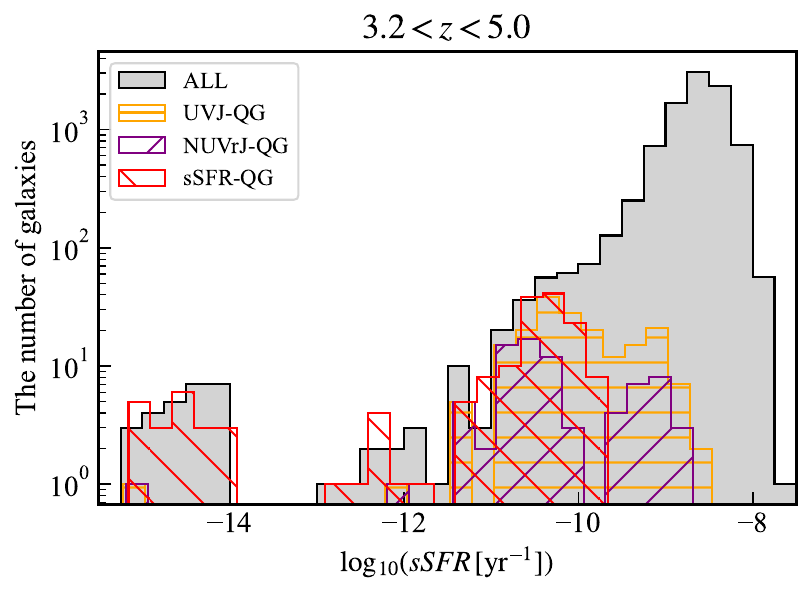}
    \end{minipage}
    \caption{$sSFR$ distribution of galaxies at $1<z<5$. The color selection methods is from $UVJ$ (the orange histogram) and $NUVrJ$ (the purple histogram). In the $sSFR$ selection (the red histogram), we utilize the criteria of 1 dex below from the SFMS \citep{Popesso2023}. All galaxies (the black histogram) are from our sample selection (Section \ref{subsec:follow}).}
    \label{fig:sSFR}
\end{figure*}

Figure \ref{fig:sSFR} shows the $sSFR$ distributions of QGs from $UVJ$ or $NUVrJ$ color selection and $sSFR$ selection method. We utilize the GALEX $NUV$ \citep{Martin2005}, recalibrated $U, V$ \citep{MazApellaniz2006}, the Supreme-Cam $r^+$ \citep{SC2002}, the 2MASS $J$ bands \citep{2MASS} to compute the color of each galaxy based on the best-fit SED from \texttt{MIZUKI} \citep{Tanaka_2015}. This illustrates that QGs selected in three different methods have a different $sSFR$ boundary. $UVJ$-QGs have higher $sSFR$ than $NUVrJ$-QGs and $sSFR$-selected QGs.

We also conduct the same statistical analysis described in Section \ref{sec:obs} with QGs selected with $UVJ$ and $NUVrJ$ color and confirm that our results remain consistent for 3rd and 10th NN density estimations. Thus, the trend of quiescent fraction as a function of environment and stellar mass do not depend on the $sSFR$ boundary of QG selection method. We note that the entire quiescent fraction is higher if $UVJ$ color selection is utilized since more galaxies are selected as QGs. In addition, there is almost no QGs left when we apply the $NUVrJ$ color criterion at $z>4$ due to the low abundance of old QGs.

\section{Example of galaxy groups at high redshifts} \label{sec:ApeB}
\begin{figure*}[tb]
    \centering
    \begin{minipage}{0.3\textwidth}
        \centering
        \includegraphics[width=0.9\textwidth]{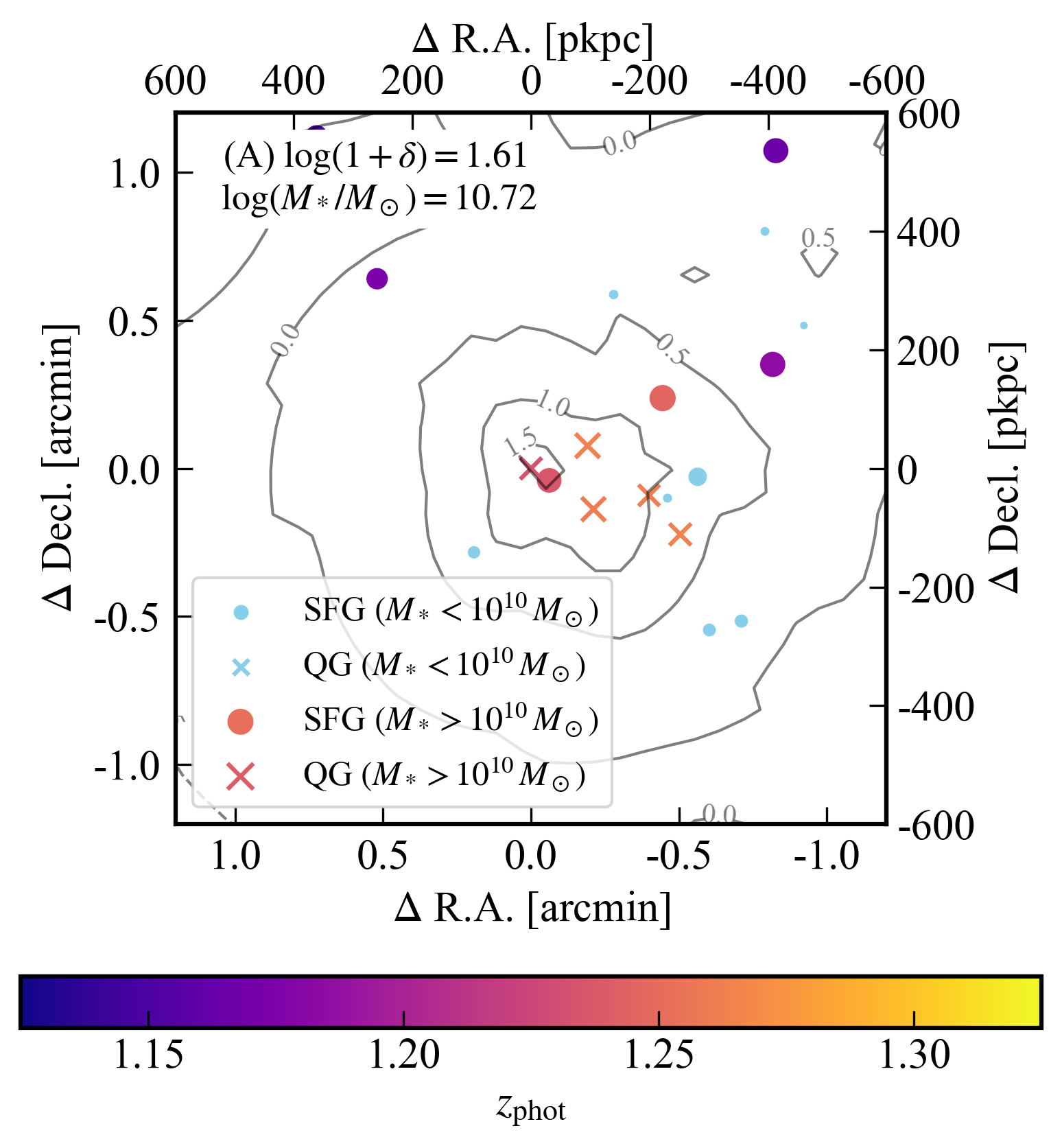}
    \end{minipage}
    \begin{minipage}{0.3\textwidth}
        \centering
        \includegraphics[width=0.9\textwidth]{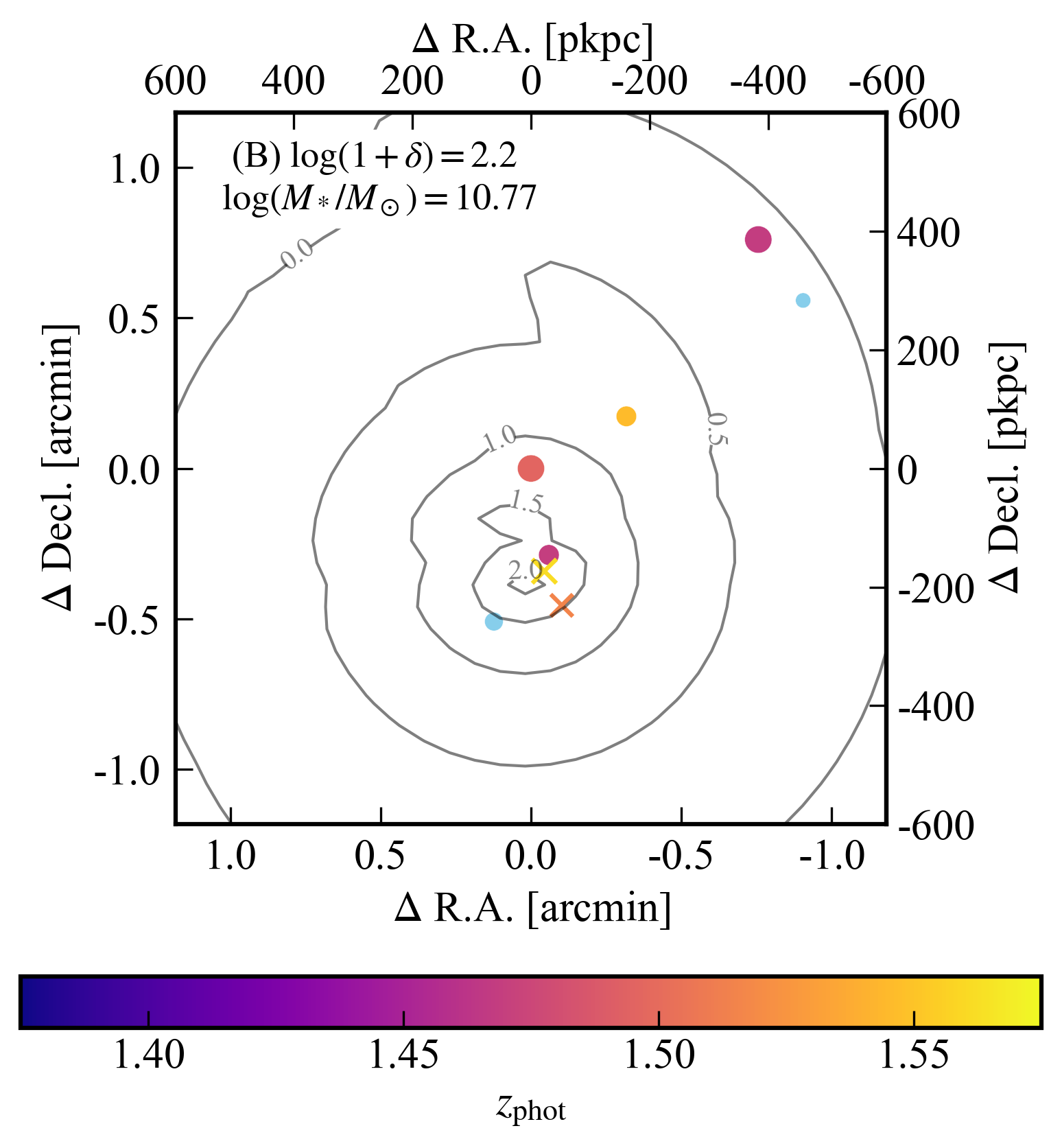}
    \end{minipage}
    \begin{minipage}{0.3\textwidth}
        \centering
        \includegraphics[width=0.9\textwidth]{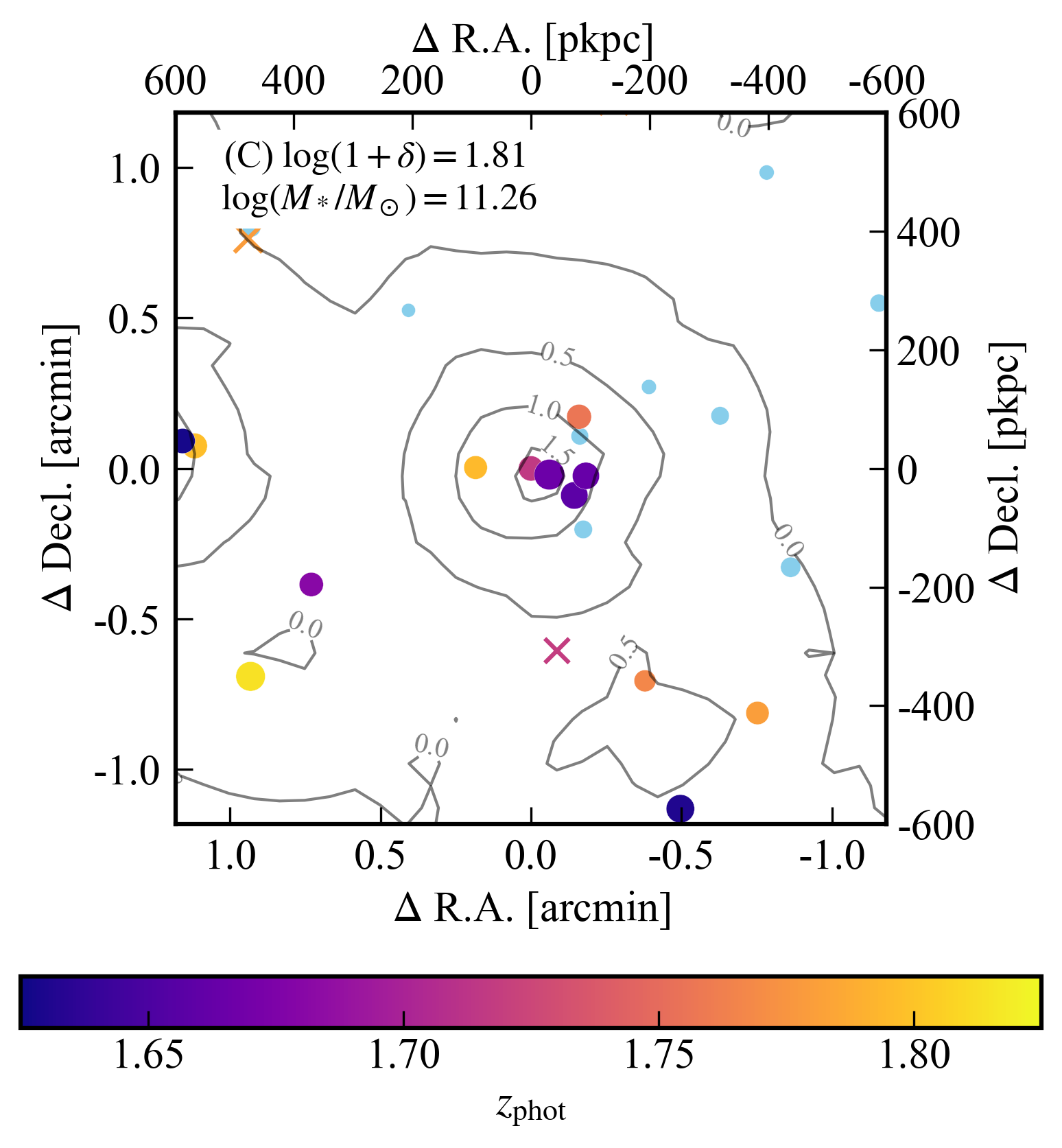}
    \end{minipage}
    \begin{minipage}{0.3\textwidth}
        \centering
        \includegraphics[width=0.9\textwidth]{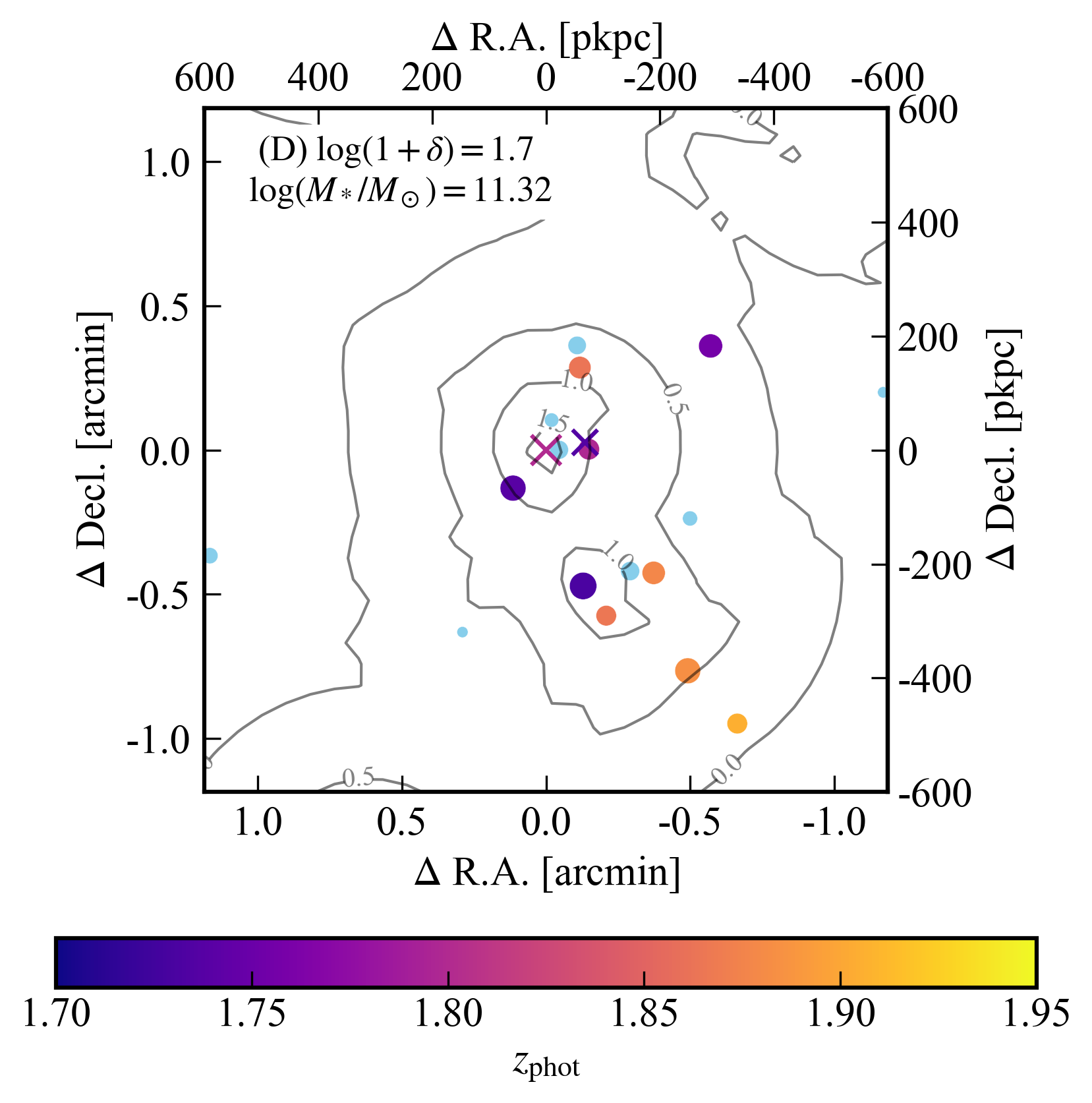}
    \end{minipage}
    \begin{minipage}{0.3\textwidth}
        \centering
        \includegraphics[width=0.9\textwidth]{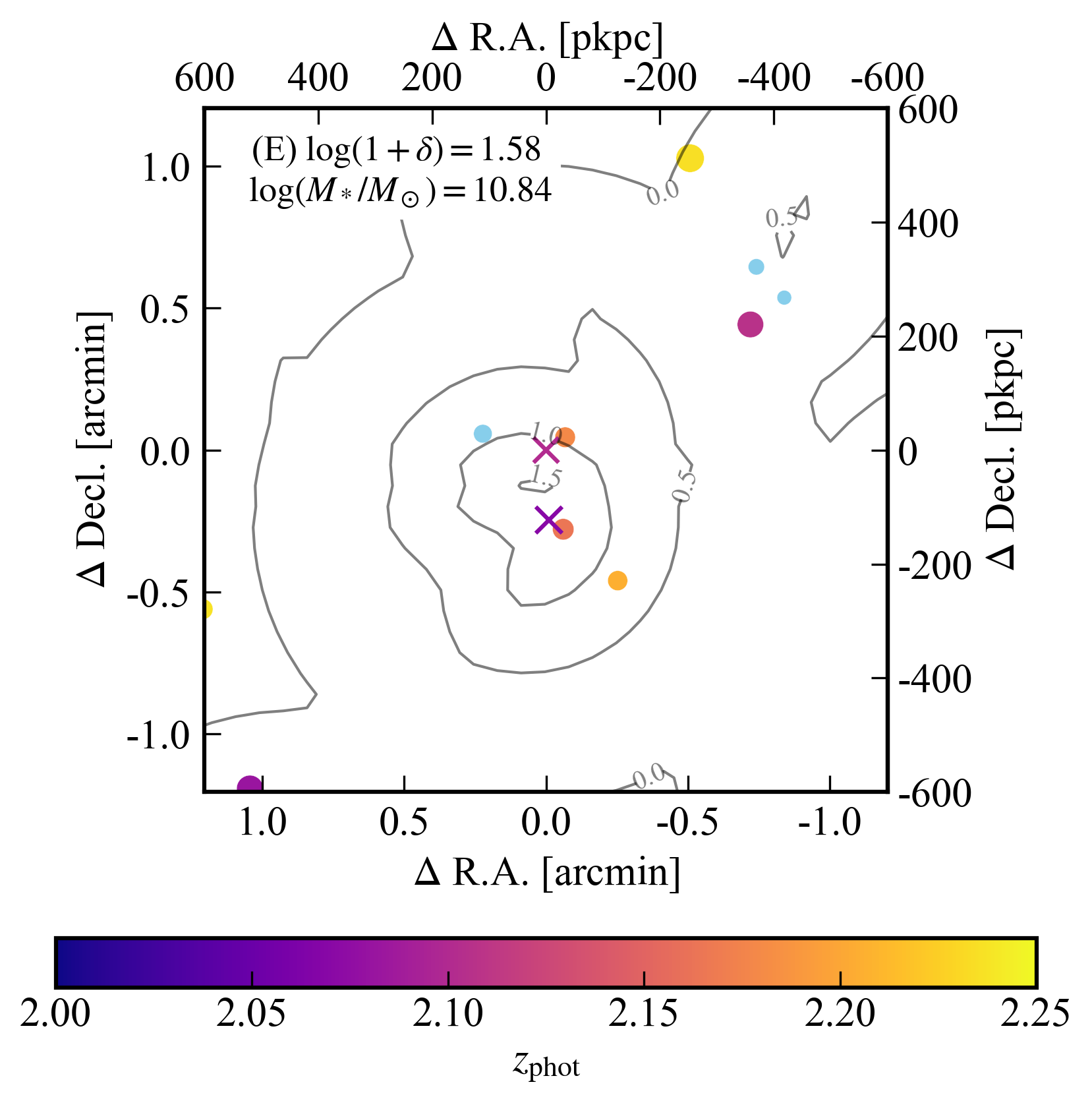}
    \end{minipage}
    \begin{minipage}{0.3\textwidth}
        \centering
        \includegraphics[width=0.9\textwidth]{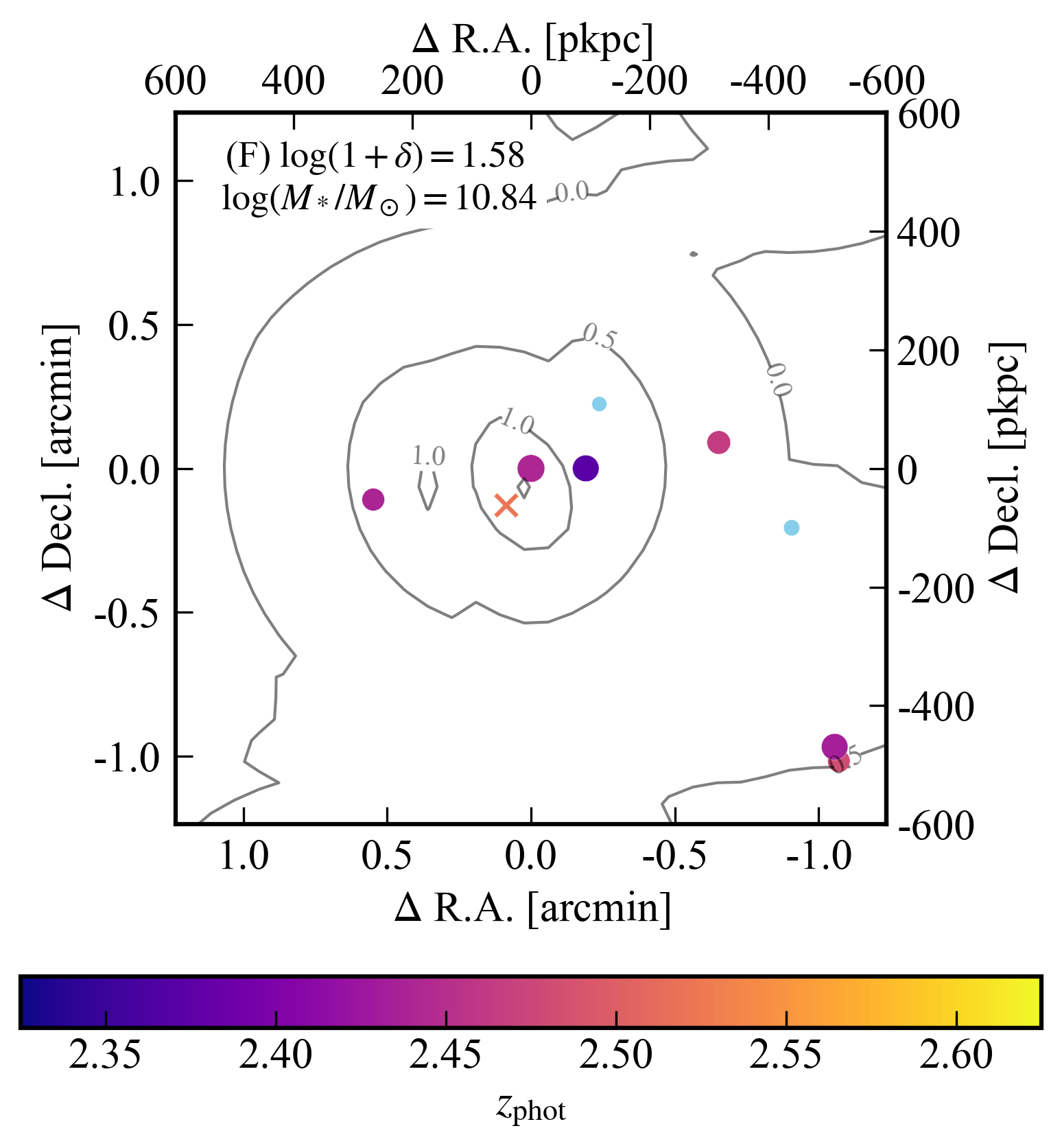}
    \end{minipage}
    \begin{minipage}{0.3\textwidth}
        \centering
        \includegraphics[width=0.9\textwidth]{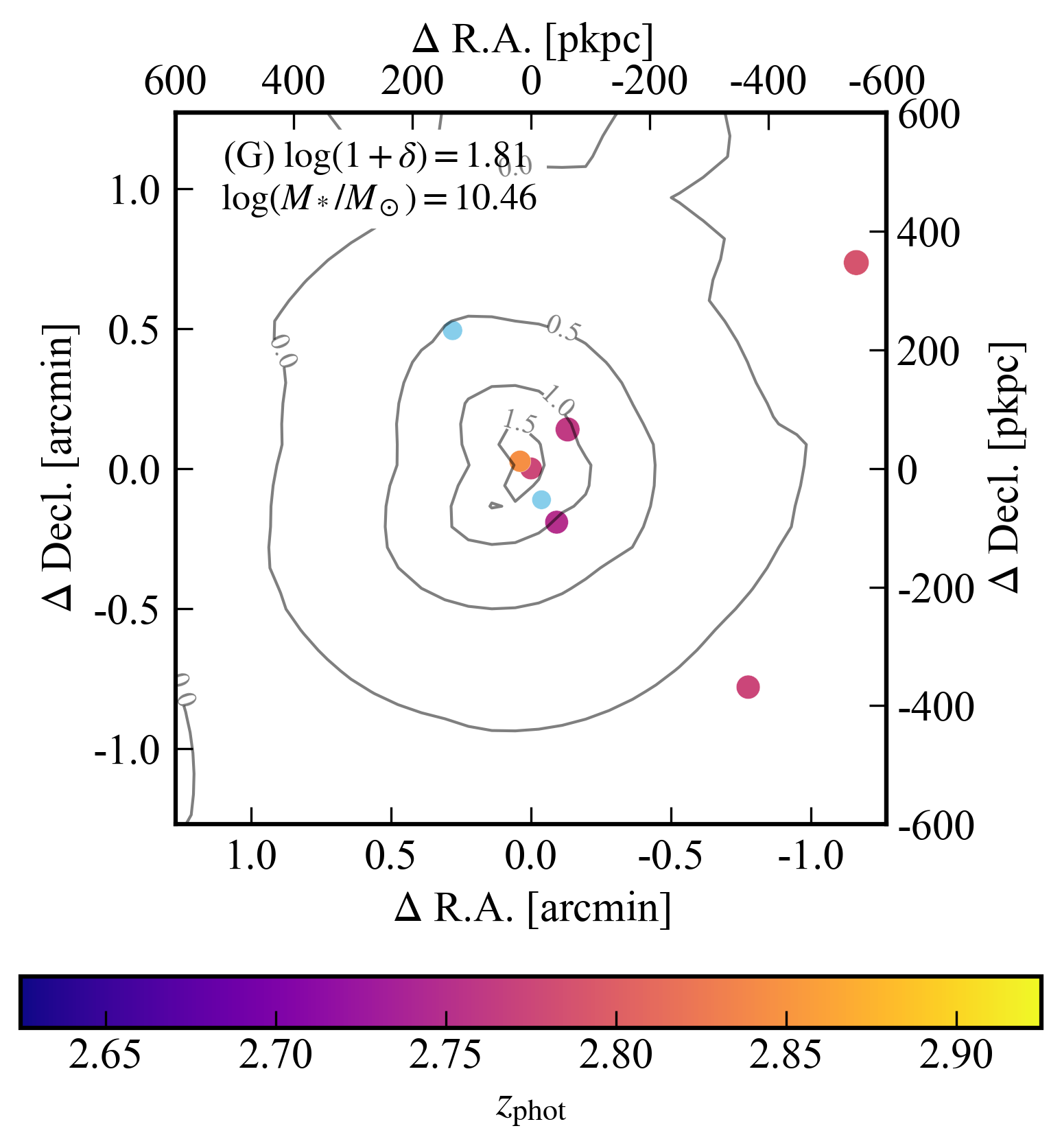}
    \end{minipage}
    \begin{minipage}{0.3\textwidth}
        \centering
        \includegraphics[width=0.9\textwidth]{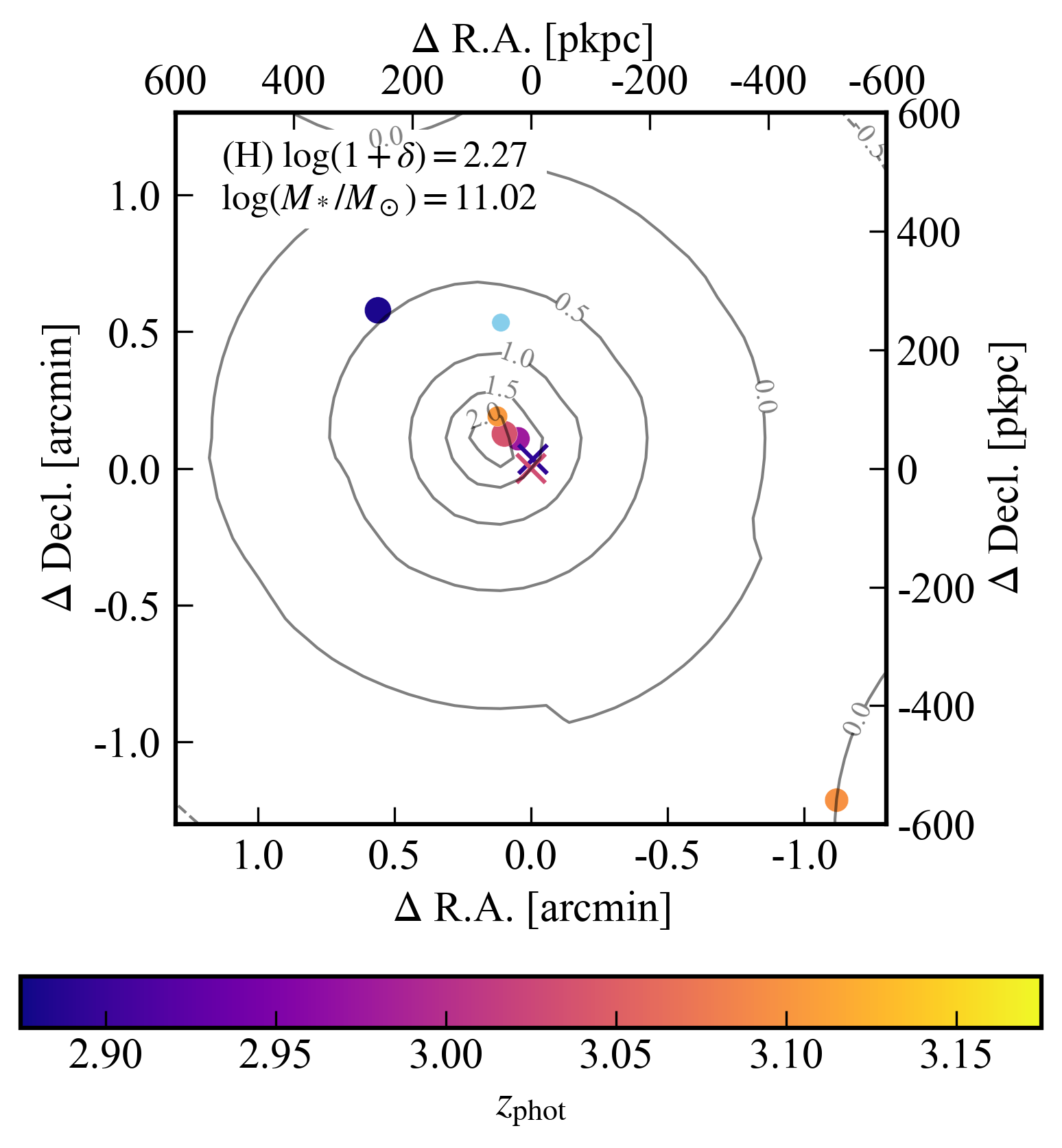}
    \end{minipage}
    \begin{minipage}{0.3\textwidth}
        \centering
        \includegraphics[width=0.9\textwidth]{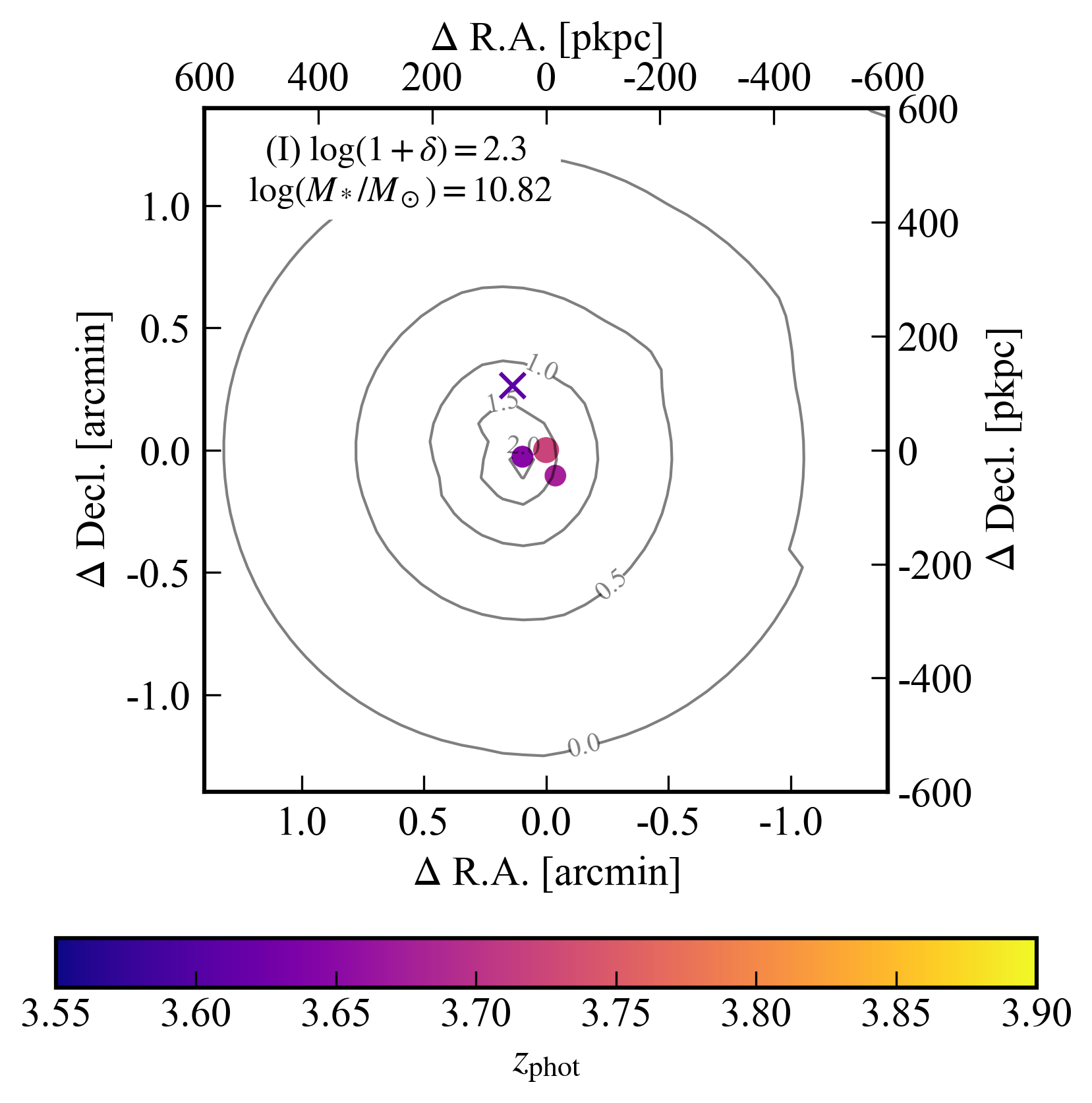}
    \end{minipage}
    \begin{minipage}{0.3\textwidth}
        \centering
        \includegraphics[width=0.9\textwidth]{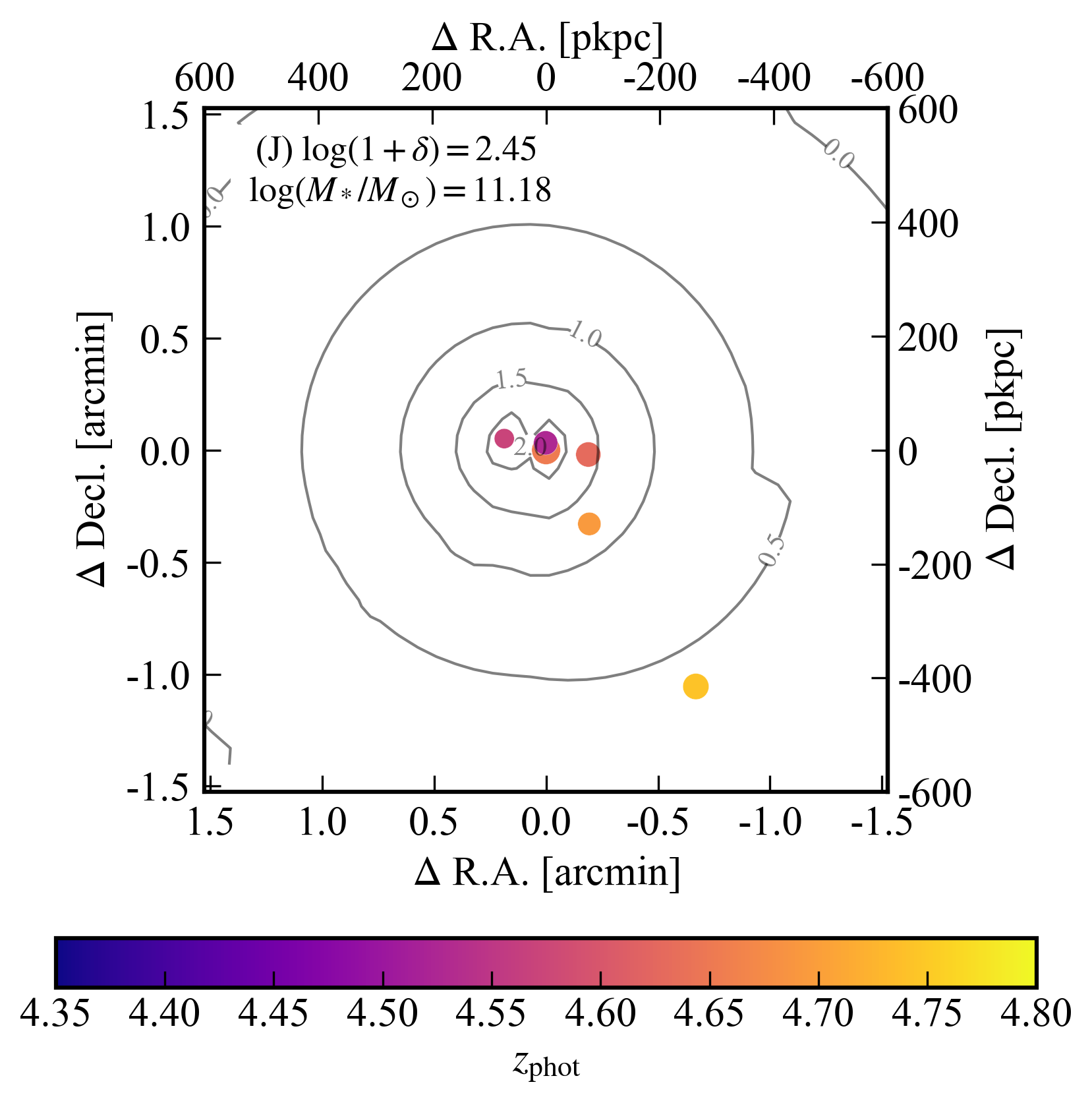}
    \end{minipage}
    \caption{Distribution of galaxies in groups at $1 < z < 5$. The gray contours show the galaxy overdensity value ($\log{(1 + \delta)}$). The filled circles and crosses indicate the positions of QGs and SFGs, respectively, color-coded by their photometric redshifts and with sizes scaled by their stellar masses. The sky-blue markers indicate galaxies with $M_*<10^{10}\,M_\odot$ that satisfy the mass completeness limit defined in Section \ref{subsec:follow} (i.e., galaxies not used for the environment measurement). Panel (J) corresponds to the group described in Section \ref{sec:453}, and panel (F) is located within the proto-cluster at $z=2.47$ \citep{Casey2015}.}
    \label{fig:groups}
\end{figure*}

Figure \ref{fig:groups} shows some examples of galaxy groups ($\log{(1+\delta)}>1$) at $1<z<5$. The gray contour represents the $\log{(1+\delta)}$, and the filled circles and crosses show the distribution of SFGs and QGs, respectively. The final panel (panel J) shows the same group discussed in Section \ref{sec:453}, but its central QG (COSMOS-1047519) is not satisfied the QG criteria when constrained solely by the photometric redshift catalog. However, QG solutions exist within the uncertainty of the photometric redshift, and the galaxy is finally classified as a QG based on constraints from the spectroscopic redshift and the rest-frame optical spectrum \citep{Kakimoto2024}. Each panel also lists the group overdensity value and the stellar mass of the central galaxy, defined as the most massive galaxy within the overdensity.


\bibliography{sample631}{}
\bibliographystyle{aasjournal}



\end{document}